# A Taxonomy of Snow Crystal Growth Behaviors: 1. Using c-axis Ice Needles as Seed Crystals

Kenneth G. Libbrecht

Department of Physics, California Institute of Technology
Pasadena, California 91125, kgl@caltech.edu

**Abstract. I describe a new approach to the classification of snow crystal morphologies that focuses on the most common growth behaviors that appear in normal air under conditions of constant applied temperature and water-vapor supersaturation. The resulting morphological structures are generally robust with respect to small environmental changes and thus should be especially amenable to computational modeling. Because spontaneous structure formation depends on initial conditions, the choice of seed crystal can be an important consideration, and I have found that slender c-axis ice needles provide an exceptionally good starting point for this series of investigations. A sharp needle tip exposes a single basal surface that often simplifies subsequent morphological development, and the absence of a nearby substrate allows for the exploration of a broad range of supersaturations with well-controlled boundary conditions. The overarching goal of this endeavor is to facilitate detailed quantitative comparisons between laboratory ice-growth experiments and corresponding computational models, which will should greatly improve our understanding of the ice/vapor molecular attachment kinetics as well as our ability to model diffusion-limited growth dynamics in the ice/vapor system. This specific case-study of water ice connects broadly to many areas in aqueous chemistry, cryobiology, and environmental science, while the physical principles of molecular attachment kinetics and diffusion-limited growth apply more generally to other systems in crystal growth and materials science.**

## ❄ Classification of Snow Crystal Morphologies

Soon after the development of optical magnifiers, numerous observers began documenting the remarkable diversity of natural snow crystal forms [1982Fra, 2002Wan]. For example, René Descartes made an especially detailed account of snow crystal structures in his famous *Les Météores* in 1637, including several varieties of stellar plates and the first recorded observations of capped columns [1637Des, 1982Fra, 2021Lib]. Robert Hooke sketched numerous snowflake forms in *Micrographia* [1665Hoo], and William Scoresby documented triangular snow crystals along with capped columns and numerous platelike forms during his Arctic voyages recounted in 1820 [1820Sco].

With the advent of photomicroscopy in the late 1800s, it became possible to document snow crystal morphologies in substantially greater detail, thus revealing a surprisingly large menagerie of complex growth forms [1931Ben, 1954Nak]. To provide a consistent nomenclature for discussing these structures, Ukichiro Nakaya developed the early classification system shown in Figure 1, consisting of 41 distinct forms. Nakaya's chart was later expanded by Magono and Lee to 80 categories [1966Mag] and recently expanded further by Kikuchi and Kajikawa to include 121 individual snowflake types [2013Kik, 2011Kik], as shown in Figure 2. Descriptive names have also been ascribed to many growth forms, and Figure 3 shows a selection of names that are in relatively common use [2006Lib1]. It should be noted that none of these charts provides a complete listing, and some snow crystals have more than one common name.

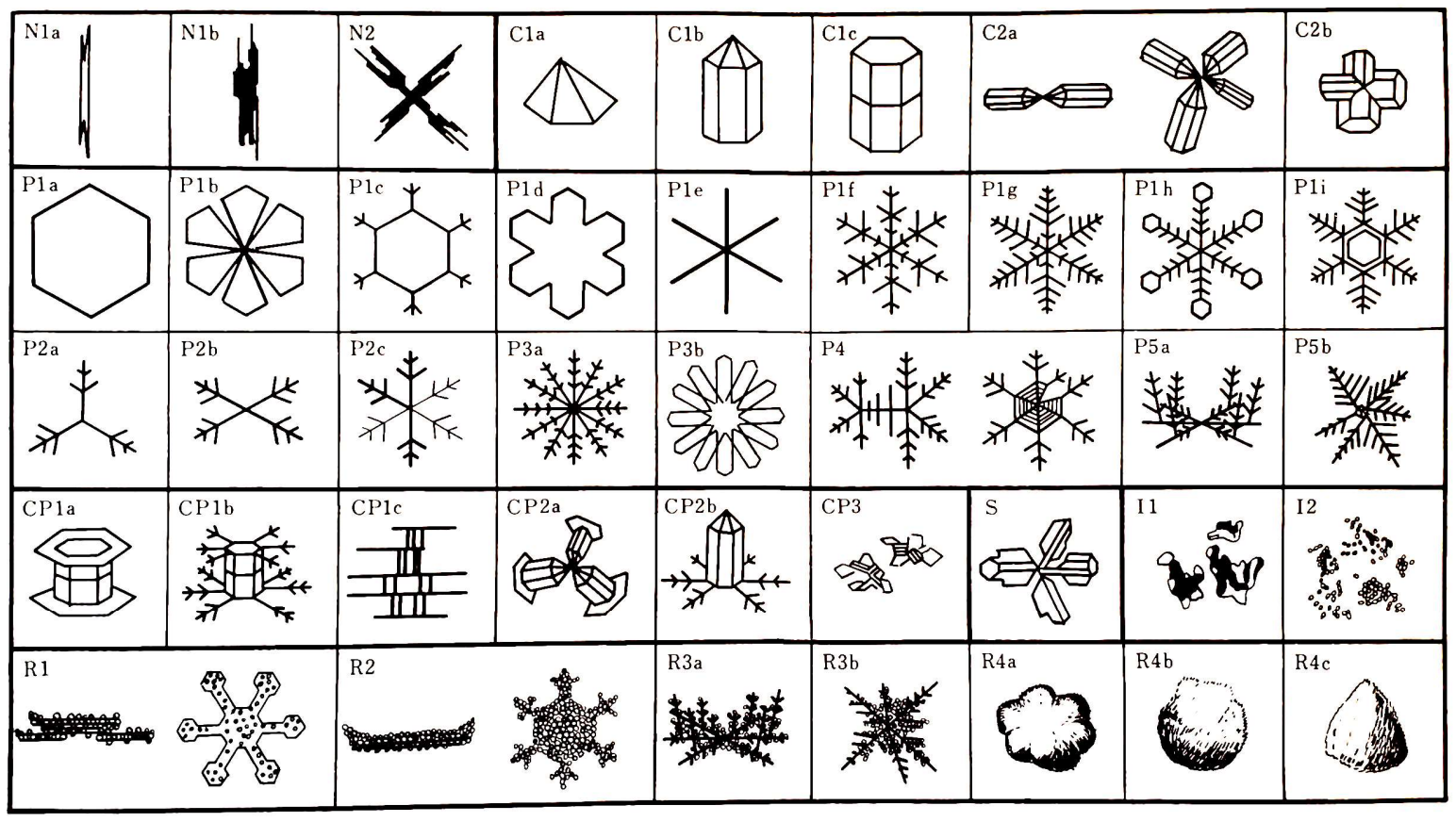

Figure 1: (left) The first morphological classification system for natural snow crystals, developed by Ukichiro Nakaya in the 1930s [1954Nak].

Figure 2: (below) This large snow-crystal classification system contains 121 separate morphological categories [2011Kik, 2013Kik].

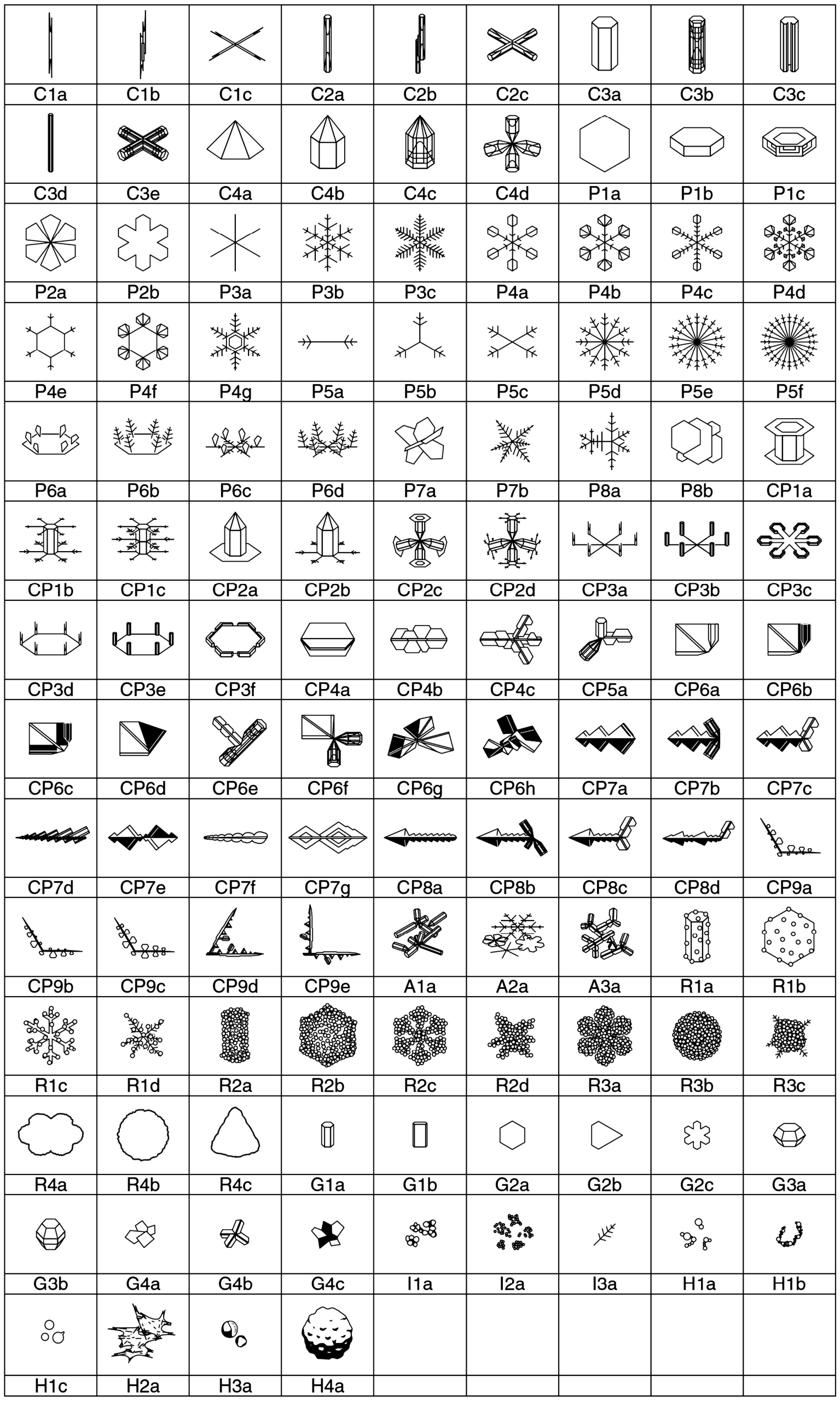

Although classification systems such as these are valuable in providing a language for discussing observations of natural snow crystals, they are not especially useful for revealing the physical dynamics that guide the formation of the different structures. Because the morphology of an individual snow crystal depends on the entire growth history it experienced, the number of possible permutations is extremely large, so no reasonable classification chart can encompass all observed forms. If our overarching goal is to understand the physical origins of the observed menagerie of snow crystal morphologies, a better approach is to focus on the fundamental processes involved in ice crystal growth and how these spontaneously generate such a rich variety of complex structures.

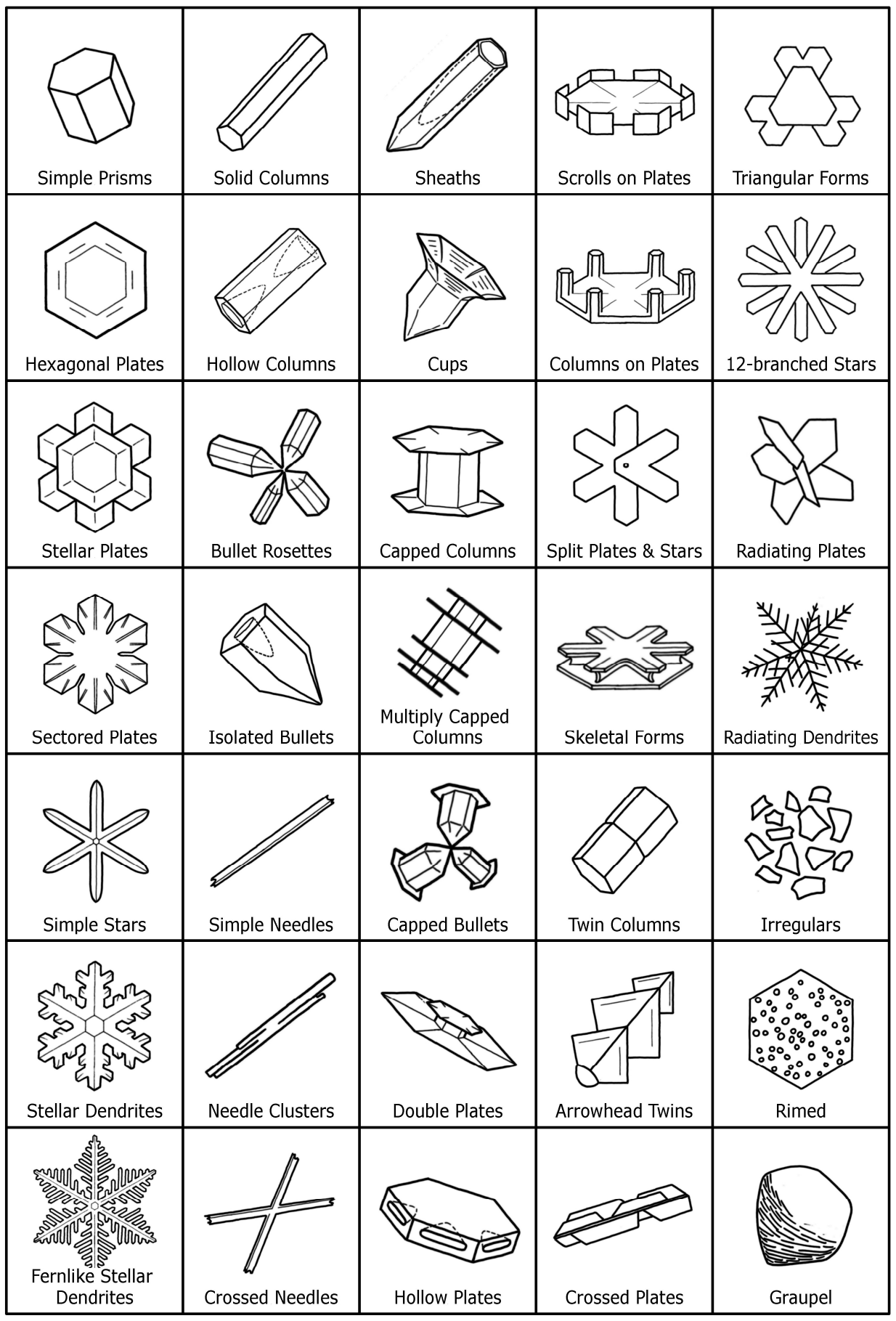

Figure 3: In addition to the alphanumeric nomenclature initiated by Nakaya, descriptive names have also been adopted for numerous snow crystal types, and this chart shows a selection that have become popular over the years [2006Lib1].

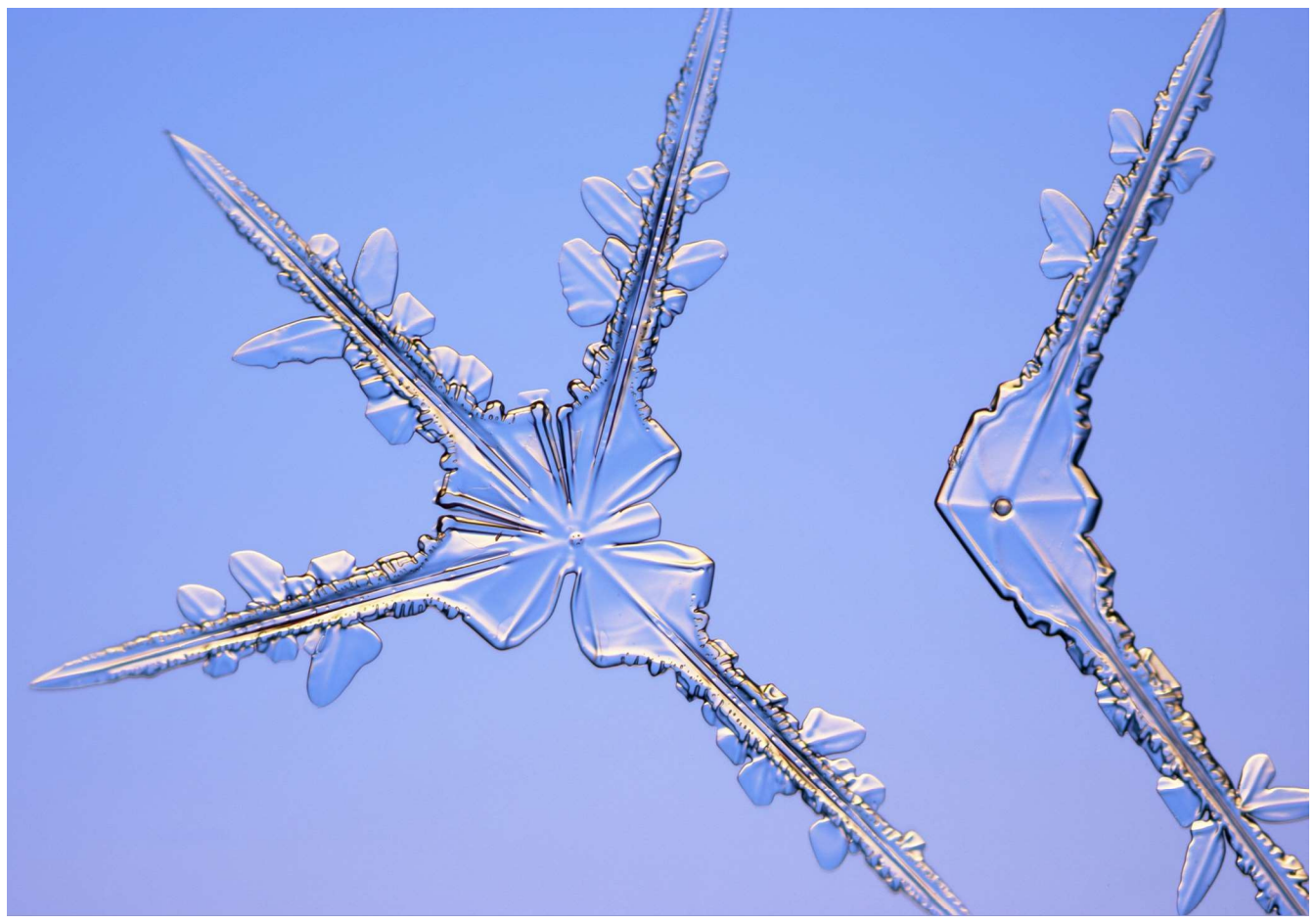

Figure 4: The above example illustrates a split star that broke apart when it fell on a collection board and was soon photographed (photo by Patricia Rasmussen [2003Lib2]). Figure 5 shows additional possible topologies for split stars.

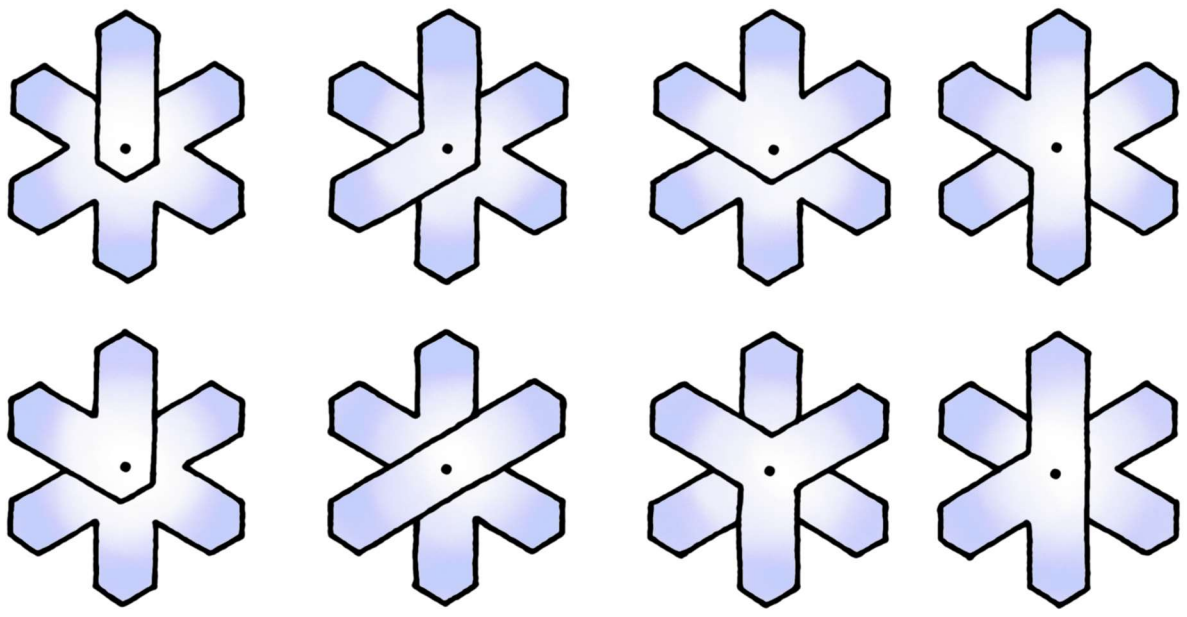

Figure 5: Several different split-star topologies can arise, depending on which branches grow out on the two capped-column plates, as larger branches on one plate stunt the growth of corresponding branches on the other plate.

## An Example – Split Stars

To better see why classification systems are not well suited for describing snow crystal growth dynamics, consider the case of split stars shown in Figures 4 and 5. The formation of these crystals begins with a simple prism that soon develops two end plates, essentially a capped column where the column is short, so the two end plates grow quite close to one another. If the hexagonal tips subsequently develop branches, corresponding branches on the two plates are so closely spaced that they interfere with one another's development. As a result, one of each pair quickly overshadows its sibling in each of the six cases. Depending on which branches dominate on which plane, several different split-star topologies are possible, as illustrated in Figure 5.

Comparing Figure 5 with Figure 2, we see that the types P5a, P5b, and P5c in Figure 2 describe some of the possibilities for broken split stars, but additional variations are not included in this classification scheme. Moreover, the P3 and P4 series do not distinguish unbroken split stars from single-plane stars, even though these are topologically different morphologies. The Kikuchi-Kajikawa classification system could be expanded further to include the full spectrum of split stars and split-star fragments, but a rather large number of specific cases must be added to describe all the different permutations of a relatively simple growth process. Not only does the classification system increase rapidly in size to include additional possible growth outcomes, it also tends to obscure the essential physics underlying the different structures.

Our take-away conclusion from this example is that there is an inevitable degree of futility in any attempt to classify and catalog all the different forms that can develop during the growth of natural snow crystals. Clearly some form of nomenclature is necessary when speaking or writing about natural snow crystals, but it provides little guidance for understanding how these complex structures emerge as ice grows from water vapor in air.

## The Nakaya Diagram

In addition to developing the first classification system for natural snow crystals, Nakaya also made pioneering observations of laboratory-grown specimens, and these quickly provided important clues for understanding the physics underlying snow-crystal growth. Plotting the basic crystal morphology as a function of air temperature and water-vapor supersaturation yielded the result shown in Figure 6, now known as the *Nakaya diagram*. These observations were soon confirmed by other researchers, yielding the similar diagrams shown in Figure 7. The cartoon version in Figure 8 is especially useful for expressing the essential features of the Nakaya diagram to a non-scientific audience.

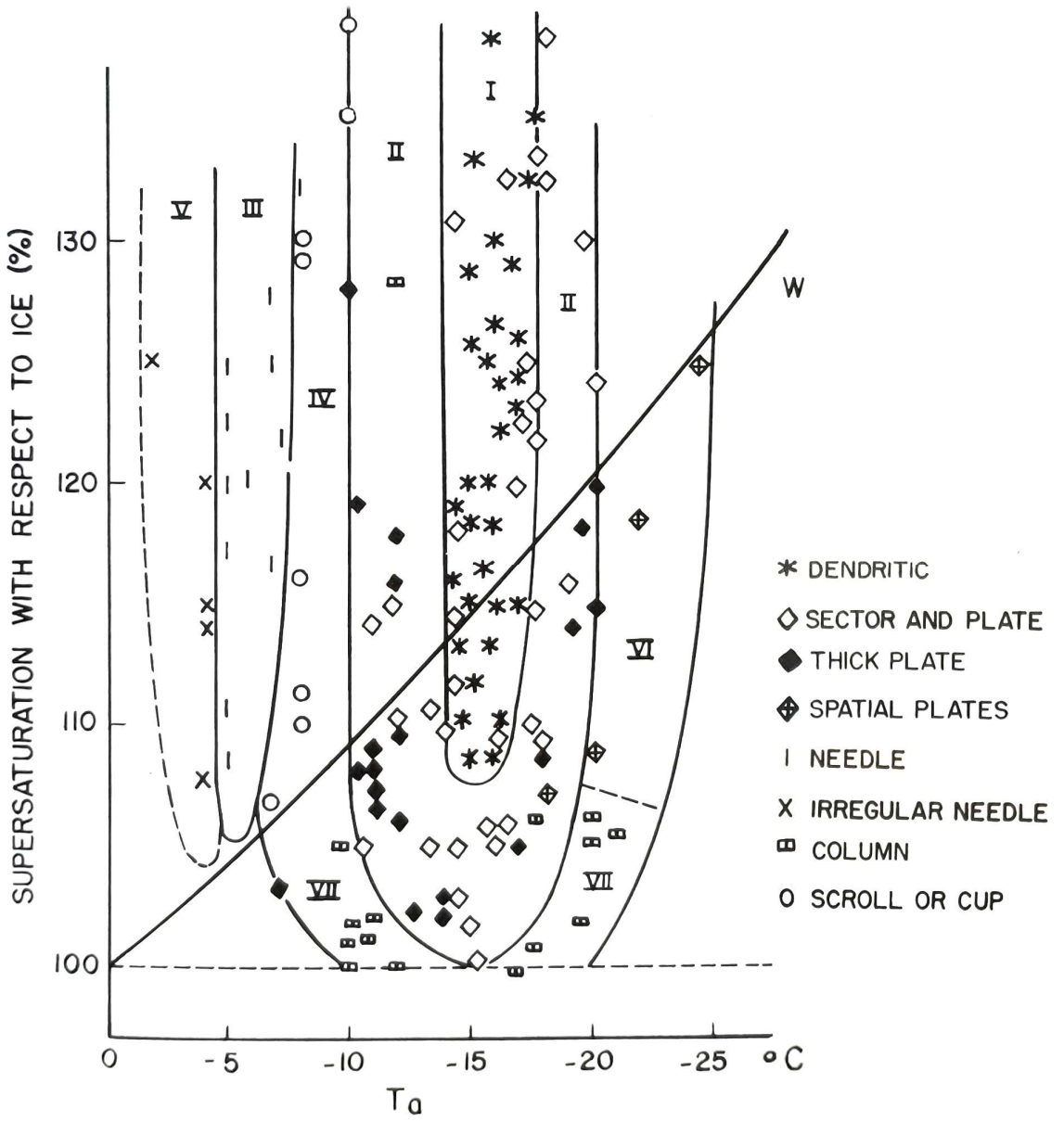

Figure 6: The first *Nakaya diagram* plotting snow crystal growth morphology as a function of temperature and water-vapor supersaturation [1954Nak].

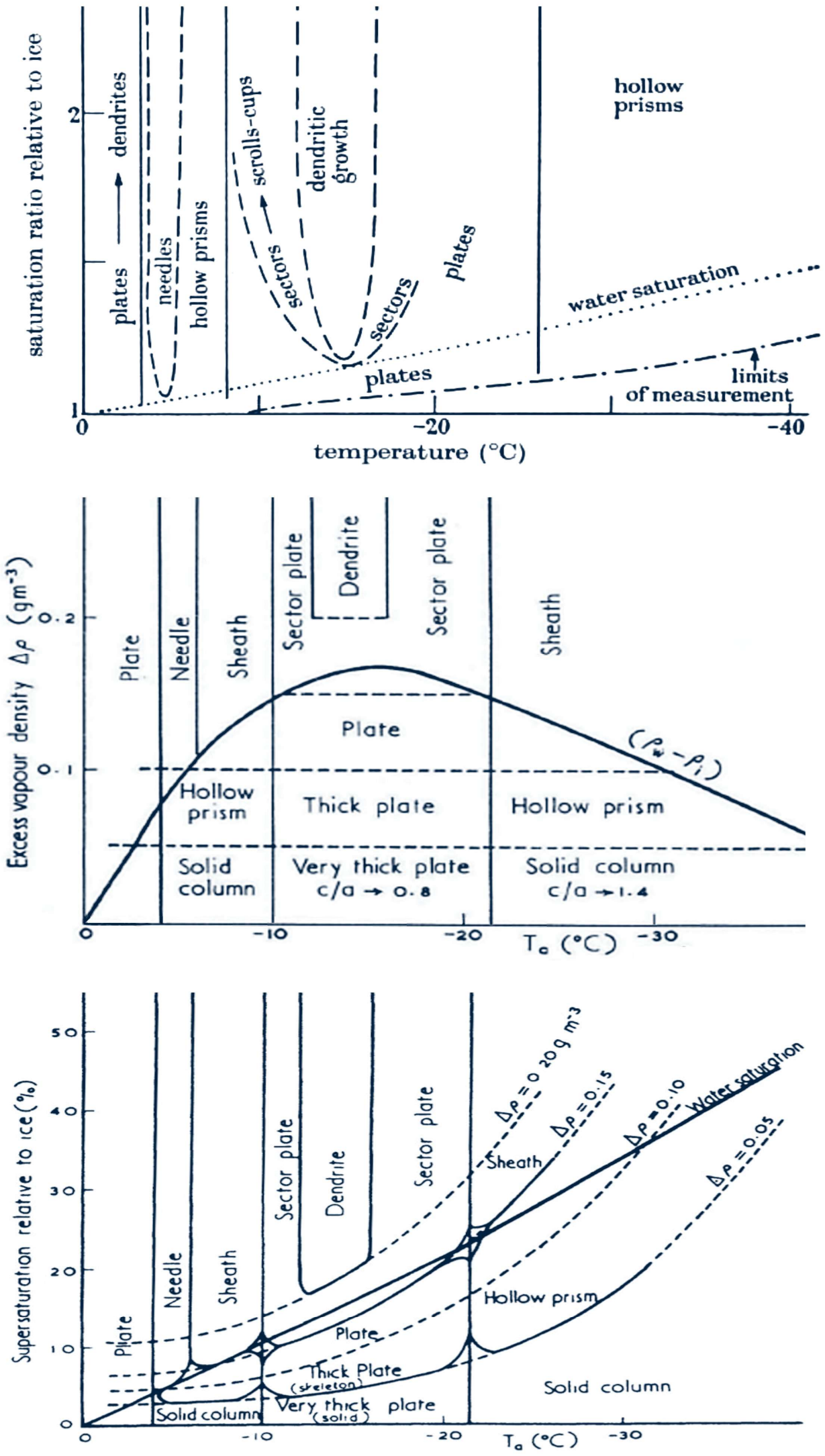

Figure 7: Additional published versions of the Nayaka diagram described in: [1958Hal] (top), [1961Kob] (middle), and [1990Yok] (bottom).

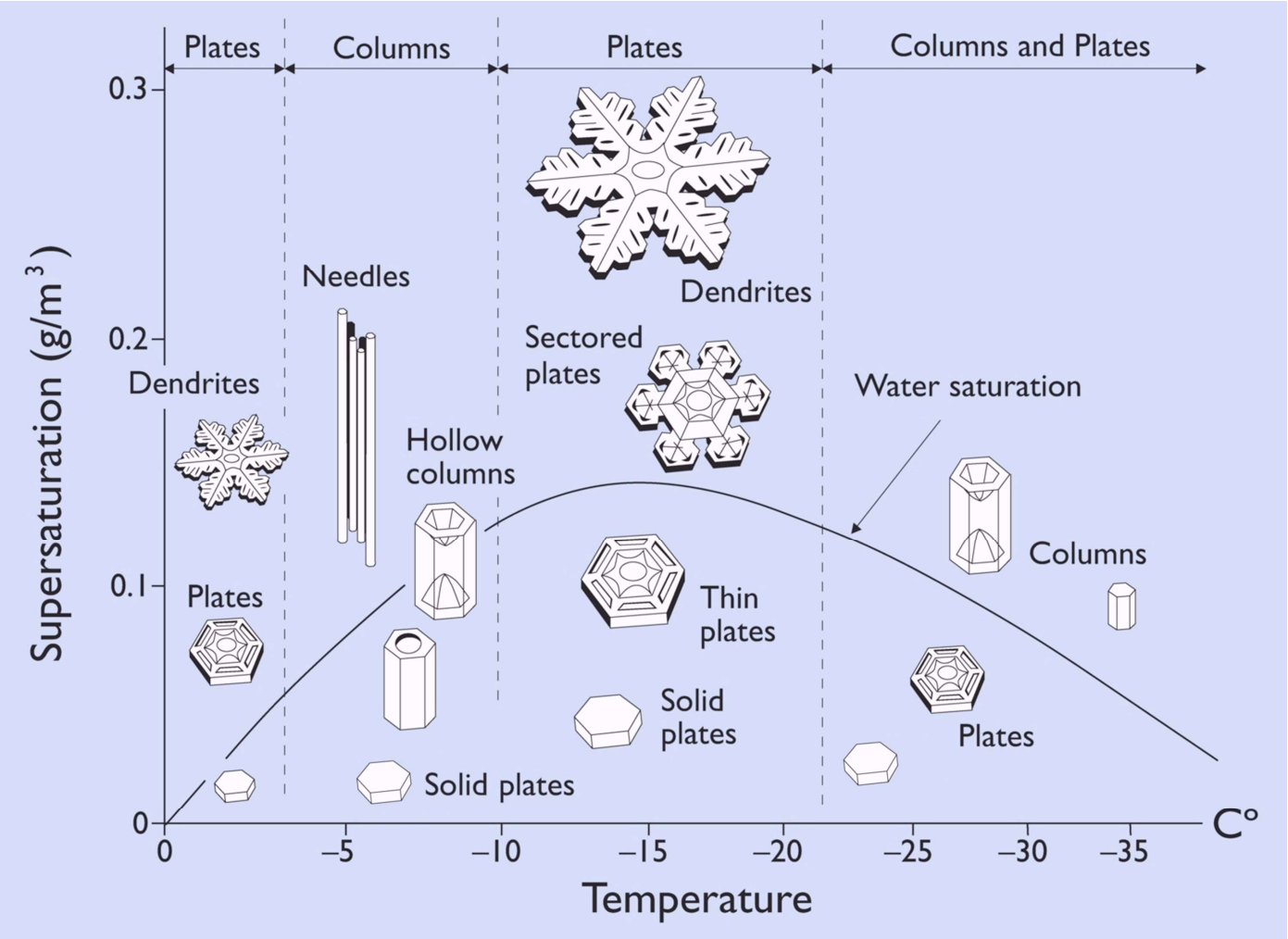

Figure 8: This cartoon version of the Nayaka diagram shows the key variations in snow-crystal morphology with temperature and supersaturation.

One important feature of the Nakaya diagram is that it describes synthetic snow crystals grown under constant environmental conditions, as opposed to the varying conditions experienced by natural snow crystals in the atmosphere. These laboratory results nicely explain numerous features seen in falling snow, with the capped column being a prominent example. As a winter cloud rises and cools, it is relatively common for a cloud droplet to freeze and commence growing at temperatures around -6 C, developing first into a columnar form. As the air mass continues to rise and cool, the growing crystal may soon experience temperatures closer to -14 C, yielding a pair of plates growing on the ends of the column – a capped column.

A warming temperature trend can yield columns growing on plates (see Figure 3), but snowfalls are usually caused by clouds cooling rather than warming, so capped columns are much more common than columns on plates. This example shows how the Nakaya diagram was a substantial advance in understanding snow crystal growth, as it provides at least some qualitative understanding of the different morphological structures included in the different classification schemes.

While the organization of growth behaviors seen in the Nakaya diagram helps explain the rich menagerie of snow crystal types, the origin of this organization requires a deeper dive into the underlying molecular physics of crystal growth. This is a nontrivial subject about which quite a lot has been written over the years [1971Mas, 1987Kob, 2001Nel, 2017Lib, 2019Lib, 2021Lib], with [2021Lib] providing the most comprehensive review to date. The basic physics of diffusion-limited growth, notably the Mullins-Sekerka instability [1964Mul], is responsible for the formation of complex dendritic structures, while the molecular attachment kinetics produces the transitions between platelike and columnar growth as function of temperature [2019Lib1, 2020Lib3]. While our recent model of the attachment kinetics provides a compelling semi-quantitative explanation of the overall temperature structure seen in the Nakaya diagram [2019Lib2, 2020Lib1, 2020Lib2], the overall state of our understanding of the molecular dynamics of ice crystal growth from water vapor remains quite rudimentary.

## ❄ Computational Modeling

The prospects for attaining a much fuller comprehension of snow crystal growth dynamics are rapidly improving with the availability of modern computational modeling techniques. At the smallest physical scales, molecular-dynamics simulations are providing new insights relating to the structure and growth dynamics at the ice/vapor interface, including surface premelting [1997Fur, 2004Ike, 2016Ben, 2020Llo] and the direct calculation of nucleation parameters like the terrace step energies as a function of temperature [2012Fro, 2019Ben, 2020Llo]. These studies tie directly to experimental measurements of ice growth rates, particularly those focusing on quantitative determinations of the ice/vapor attachment kinetics [2013Lib, 2021Lib, 2020Lib1]. The ice/vapor nucleation process seems especially amenable to further investigation using these methods, as the terrace step energies naturally connect molecular processes on the ice surface to crystal growth rates on macroscopic scales [2021Lib].

On larger scales, computational studies of diffusion-limited ice growth from water vapor have already reproduced the basic features of snow crystal structure formation like faceting and branching, and numerical models are beginning to replicate subtler three-dimensional (3D) features like ridges on branches [2009Gra, 2013Kel, 2014Kel].

Cellular-automata methods seem to be especially well-suited for modeling a variety of snow-crystal growth phenomena, as illustrated in Figures 9 and 10. However, other computation techniques show promise as well [2012Bar, 2017Dem, 2017Dem1], and it remains to be seen how these different modeling strategies will be further developed and applied to the specific problem of snow crystal growth.

Unfortunately, all the 3D computational snow-crystal models presented in the literature to date suffer from two fundamental problems: 1) they all use physically incorrect or incomplete representations of the molecular attachment kinetics, and 2) none has been validated by quantitative comparisons with measurements of laboratory-grown snow crystals created in well-known environments. Nevertheless, as seen in Figures 9 and 10, these computational models already yield morphologies that look remarkably like sectored-plate snow crystals, including realistic manifestations of faceting, branching, and ridge formation. This suggests that computational models are clearly making rapid progress toward the goal of accurately modeling real snow crystals, plus it suggests that some prominent structural features are somewhat robust with respect to changes in the small-scale physics, depending more on the large-scale behavior of diffusion-limited growth.

Figure 11 shows one demonstration of how computational models can be compared directly with experimental data to good effect,

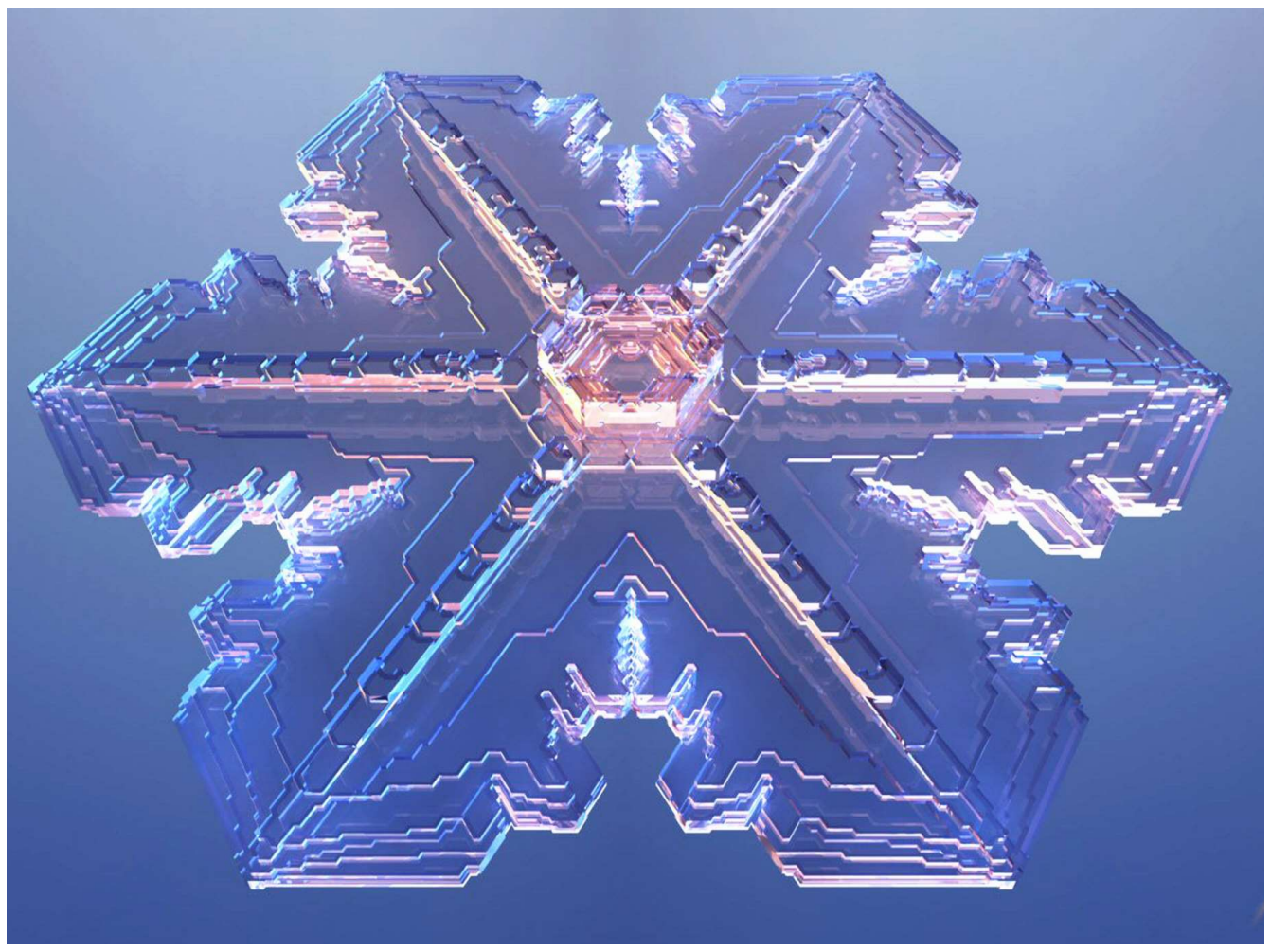

Figure 9: A Gravner-Griffeath three-dimensional snow crystal model [2009Gra] rendered by Antione Clappier, illustrating faceting, branching, and ridge-like structures on the branches.

Figure 10: (below) The growth a single snow crystal model created by Kelly and Boyer [2014Kel]. This model uses more accurate attachment kinetics than that in Figure 9, but it yielded generally similar structural features. This suggests that some morphological characteristics in snow crystal growth are rather insensitive to the details of the molecular surface physics and result mainly from diffusion-limited growth.

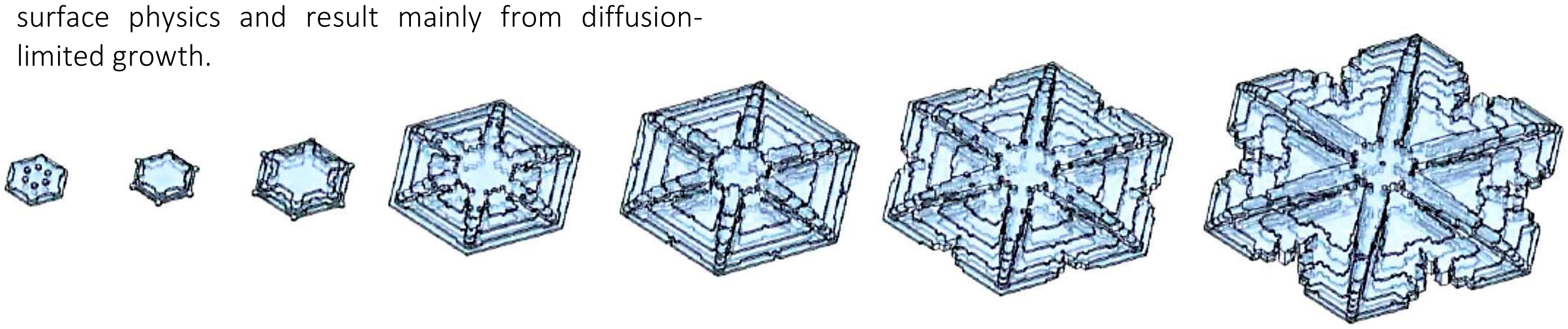

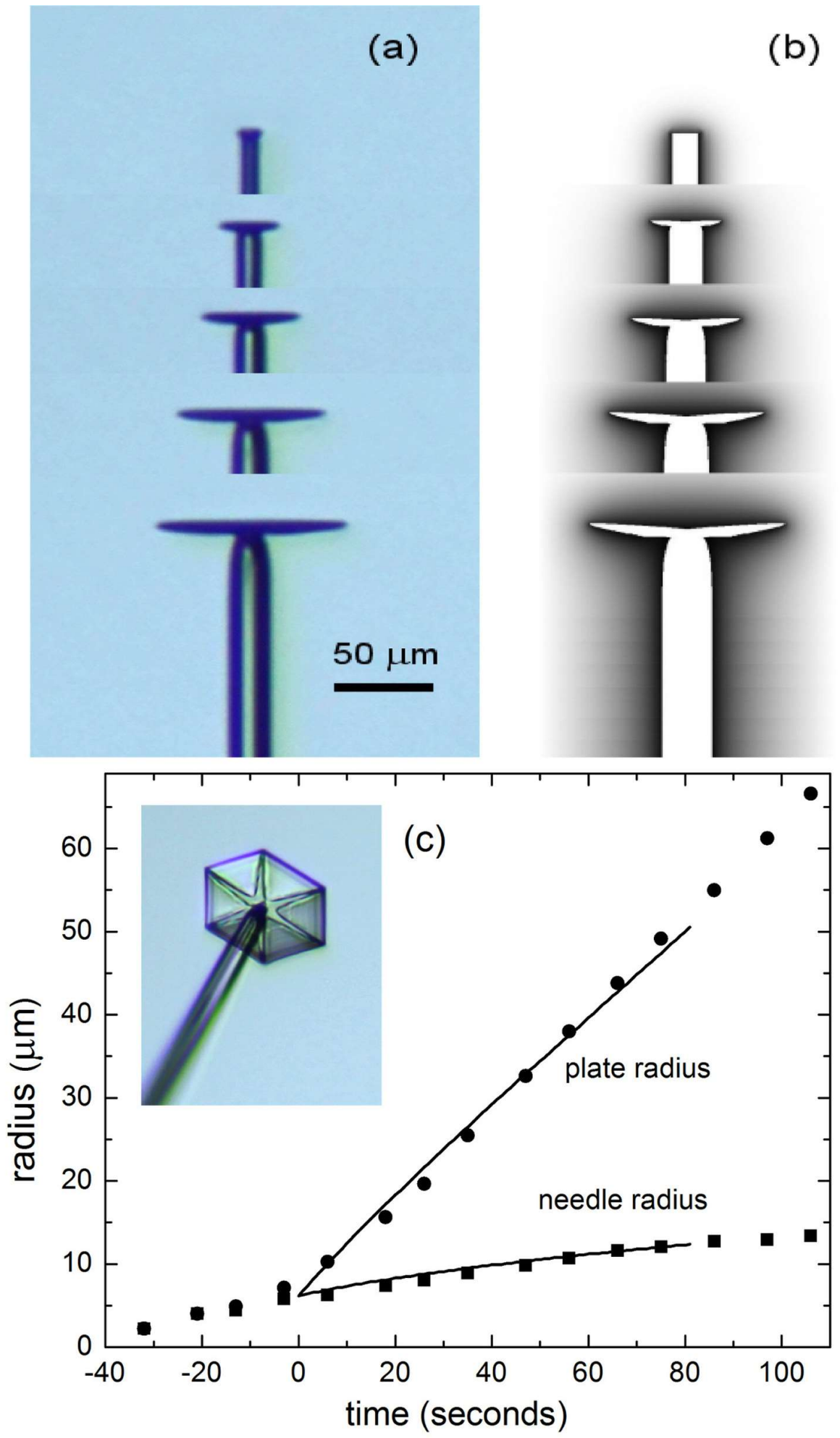

Figure 11: These figures illustrate a quantitative comparison a two-dimensional cylindrically symmetrical cellular-automat model with experimental data [2008Lib, 2013Lib1]. (a) A composite image made from five photographs shows the growth of a plate-like snow crystal on the end of an electric ice needle, viewed from the side. (b) A cylindrically symmetrical model in (r,z) space reproduces the observations, while brightness around the crystal shows the water-vapor diffusion field. (c) A quantitative comparison of experimental data (points) and the computational model (lines) shows good agreement. The inset photo shows the crystal in (a) from a different angle.

in this case modeling the growth of a platelike crystal on the tip of an ice needle [2013Lib1]. This two-dimensional (2D) cylindrically symmetrical model used a physically derived parameterization of the attachment kinetics, with the parameters adjusted (within reasonable limits) to fit the experimental data. Comparisons like this were crucial in the development of a comprehensive physical model of the attachment kinetics [2019Lib1, 2021Lib], and additional work along these lines provides a clear route to further understand how molecular processes contribute to the morphological organization seen in the Nakaya diagram.

## ❄ A Closer Look at Structure Formation

The primary goal of this paper is to initiate a series of experimental observations to facilitate detailed quantitative comparisons between laboratory measurements and computational models of snow crystal growth. Most recent 3D modeling studies have been somewhat qualitative in nature, demonstrating that they can produce realistic diffusion-limited-growth structures that resemble those seen in some snow crystal photographs. While this is an excellent initial step, achieving morphological similarity is not the same as modeling a real physical phenomenon. Further progress will certainly be expedited by making detailed comparisons with experimental data.

While computational modeling techniques have been evolving rapidly along several fronts, a substantial obstacle to validating these models is the paucity of suitable quantitative observations of laboratory snow crystals grown over a broad range of controlled conditions. To date, experimenters have simply not developed the tools needed to make observations that connect well to modeling efforts, yielding both 3D morphologies and growth rates. In an effort to ameliorate this problem, the sections below focus on how one goes about producing a suitable set of observations.

A central consideration in this research direction is the ice/vapor attachment kinetics, as this is the key element needed to create accurate models. Modeling diffusion-limited growth is generally straightforward because the underlying physics is well understood, and the slow-growing ice/vapor system allows for substantial mathematical simplifications relative to most solid/liquid systems [2021Lib]. Particle diffusion of water molecules through air is the primary player, although heat diffusion contributes small effects as well, and both are quite amenable to computational modeling. Surface energy presents a nontrivial factor in the boundary conditions, but its overall impact is relatively small in snow crystal growth, manifesting itself mainly via the Gibbs-Thomson effect [2019Lib, 2021Lib]. Importantly, one can generally assume an isotropic surface energy in snow crystal growth models with little ill effect, because any small anisotropy in the surface energy is completely dwarfed by the enormous anisotropy in the molecular attachment kinetics. This statement has not yet been fully embraced by researchers, but it is supported by substantial experimental and theoretical evidence [2012Lib2].

The molecular attachment kinetics are the most difficult aspect of snow crystal growth to model, as several molecular processes act together in subtle and poorly understood ways. Surface premelting, terrace nucleation dynamics, and surface transport effects all contribute to determining the growth rates of faceted ice surfaces. Together, these molecular-scale phenomena all factor importantly into the structure-dependent attachment kinetics that governs the formation of thin plates and hollow columnar crystals [2019Lib1, 2021Lib].

From all these considerations, it becomes clear that obtaining a physically accurate picture of snow crystal growth dynamics will require many detailed investigations over a broad range of temperatures, supersaturations, and background gas pressures. Developing a suitable quantitative model of the attachment kinetics presents a particularly difficult challenge. Adjusting *ad hoc* model parameters to fit a specific snow crystal type, especially one defined by still photographs, will simply not be sufficient to develop a comprehensive understanding of the essential physical processes involved.

## ❄ Initial Conditions and Ice Seed Crystals

Making quantitative comparisons between computational models and detailed laboratory measurements requires a careful examination of the experimental methods employed when observing snow crystal growth. One important consideration is the type of seed crystal used to begin the growth process, because growth morphology often depends on the specific initial conditions in an experiment.

For example, ice growth behaviors near -5 C in air illustrate how a change in starting conditions can lead to markedly different morphologies as a crystal develops. As demonstrated by Knight, both platelike and needlelike crystals can grow under essentially identical conditions at this temperature, presenting a remarkable juxtaposition of growth forms appearing side by side during a single experimental run [2012Kni]. Subsequent investigations showed that this phenomenon can be explained using a comprehensive model of the molecular attachment kinetics [2019Lib2], but it also shows the importance of carefully managing the initial conditions in ice growth experiments.

To this end, it is instructive to examine different types of ice-growth experiments and how the choice of seed crystal and surface boundary conditions can influence the morphological development and growth rates that are observed. There are three general categories that are most prevalent in the literature to date – ice growth on a substrate, levitated or freely falling samples, and growth on slender ice needles.

**Growth on a substrate**. I define this category as being when a seed crystal is in contact with a substrate that may influence the growth of some parts of the crystal. Examples include a small ice prism resting on a flat substrate or a seed crystal attached to some type of filamentary support. For example, Nakaya used rabbit hair to support growing snow crystals [1954Nak], while slender capillary tubes have been used in some recent experiments [1996Nel]. Small ice prisms growing on flat substrates have a long history in ice growth measurements [1972Lam, 1982Bec2, 1982Gon, 1994Gon, 2013Lib].

One pervasive concern with this class of experiments is the possibility of substrate interactions that substantially increase terrace nucleation rates, as shown in Figure 12. This problem can be mitigated by using sufficiently hydrophobic surfaces, but finding robust surface coatings for this purpose remains an ongoing challenge. Another mitigation strategy is to focus one's attention on faceted surfaces that do not intersect the substrate [2013Lib, 2015Lib3], but this is not always practical when one is interested in the full morphological development of growing crystals over a broad range of environmental conditions.

Another significant issue with substrates is the rapid nucleation of water droplets whenever the near-surface supersaturation exceeds $\sigma_{water}$, the value for liquid water with respect to ice. Water droplets strongly alter the supersaturation boundary conditions surrounding the crystal, making it quite difficult to examine ice growth behaviors on substrates with $\sigma > \sigma_{water}$.

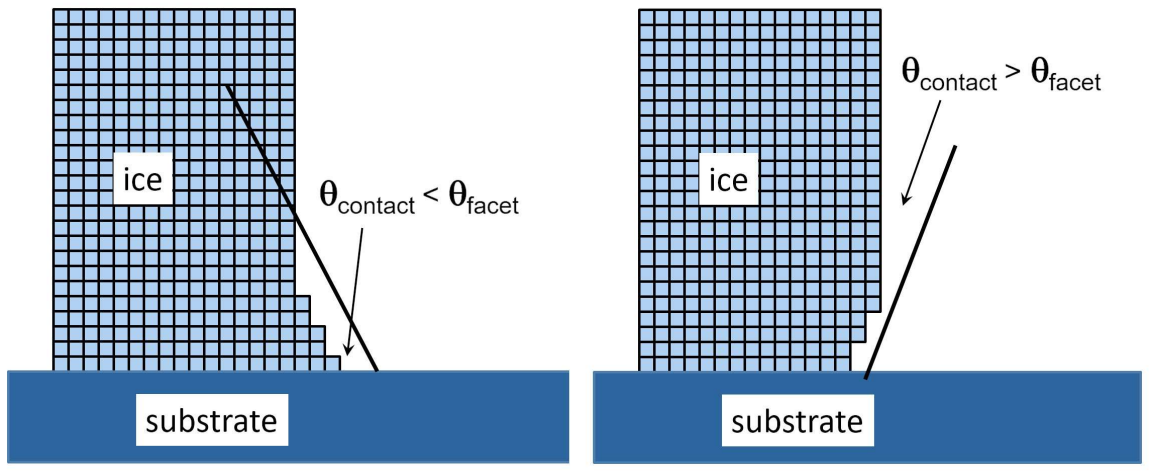

Figure 12: When an ice crystal rests on a hydrophilic surface (left), a small ice/substrate contact angle results from molecular attraction between the two substances. The resulting structure provides a source of terrace steps on the contacting facet surface, often yielding faster growth than that limited by normal terrace nucleation. This issue does not exist on a sufficiently hydrophobic surface (right). Chemical residues on the surface add additional uncertainty to the level of substrate interactions.

**Levitated or freely falling samples**. Nucleating tiny ice crystals and injecting them into a large volume of supersaturated air provides a simple method for observing large numbers of crystals [1949Sch, 1976Gon, 1987Kob, 2008Lib4], although observing the development of individual crystals over time is not possible using the free-fall method. Levitating single crystals allows an especially promising method for observing morphological development under well-controlled conditions with no substrate interactions, and recent experiments have demonstrated good control of supersaturation levels [1999Swa, 2003Bac, 2016Har].

Supporting crystals in a laminar flow chamber is another form of levitation, and a large set of crystals were observed this way by Takahashi and Fukuta [1988Tak, 1991Tak, 1999Fuk]. As with substrate-based experiments, levitation and laminar-flow techniques tend to work best when the supersaturation is no higher than $\sigma_{water}$, although careful chamber design can push beyond this limit [2016Har].

While a small, freely floating prismatic seed crystal is something of an ideal case for ice-growth experiments, it presents a symmetry issue when comparing observations with computational models, which I call the "double-plate" problem illustrated in Figure 13. In short, the presence of two basal surfaces on the seed crystal can lead to the formation of a double-plate crystal, and this form is frequently observed in 3D snow-crystal models, as illustrated in Figure 14.

The problem here is that the two sides of a double plate tend to interfere with one another's growth, often yielding split plates and split stars as described in Figure 5. This type of growth instability is typically not included in computational models, and it significantly complicates quantitative analyses. Of course, not all prisms develop into double plates, but this does occur over substantial and quite interesting regions of parameter space.

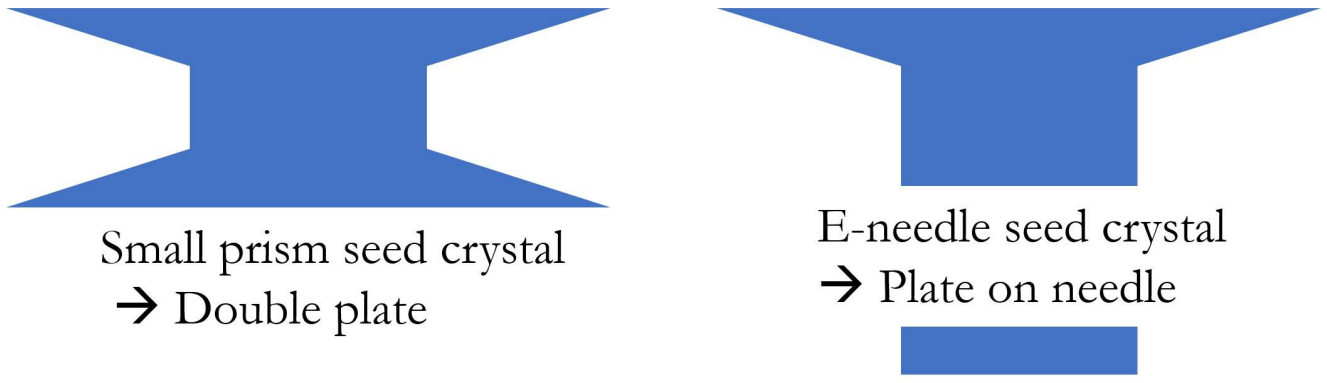

Figure 13: An illustration of the "double-plate" problem when using small, prismatic seed crystals (left). In many important growth scenarios, the presence of two basal faces on this seed crystal leads to the formation of a double-plate crystal (essentially a short capped column), and the two plates interfere with one another's subsequent growth. In contrast, the tip of an ice needle crystal (right) exposes only a single basal surface, which simplifies comparisons with computational models.

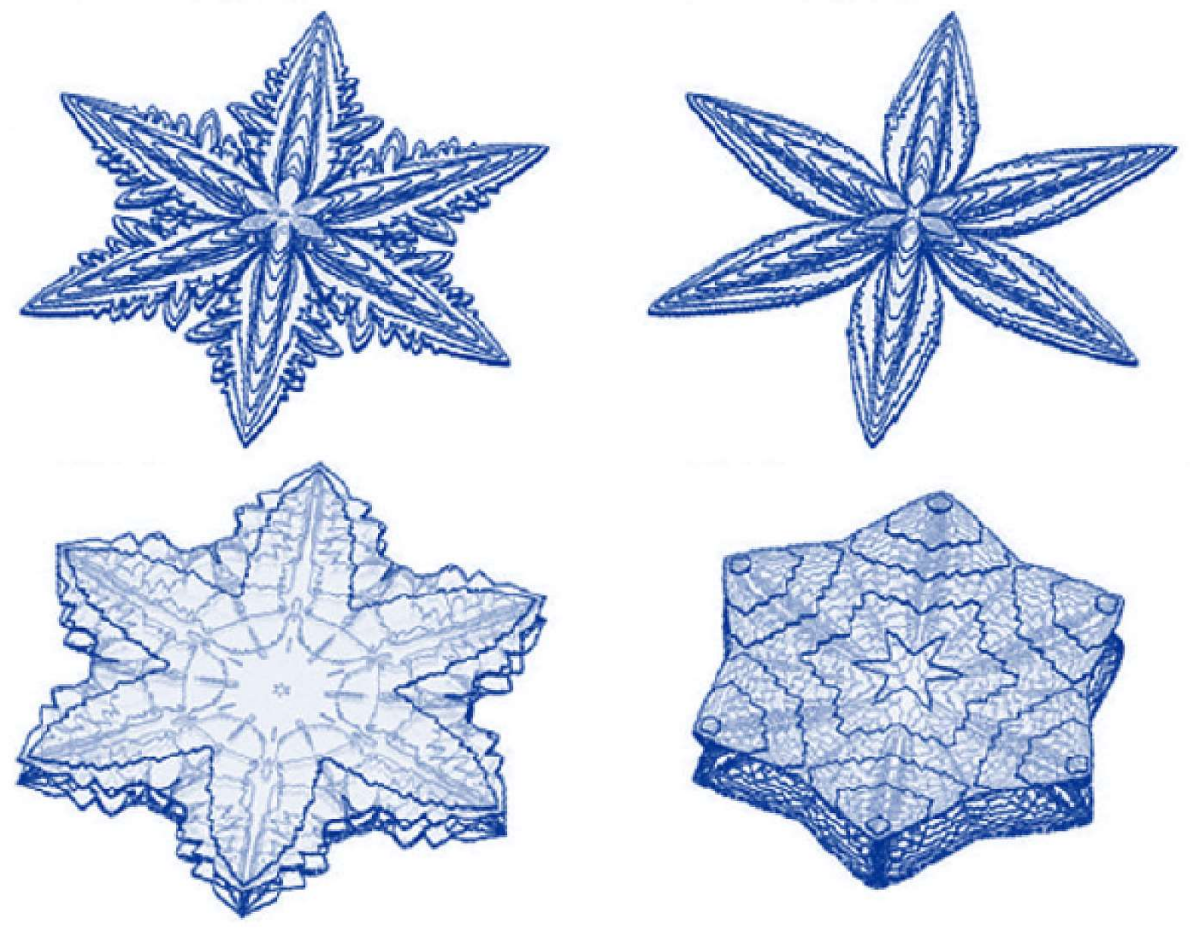

Figure 14: Beginning with a small ice prism as the seed crystal, computational models often produce double-plate crystals (the lower two example in this figure) [2014Kel]. This reflection symmetry is typically imposed in the model mathematics, while the same symmetry is often unavoidably broken by the Mullins-Sekerka instability in real crystals. Dealing with this "double-plate problem" complicates the comparison between computational models and laboratory observations.

**Growth on ice needles**. Using a slender ice needle as a seed crystal nicely avoids the double-plate problem, as the needle tip presents only a single basal facet. When plate formation is favored, only a single plate emerges from the tip, providing a relatively simply geometry for comparison with computational models. Moreover, the intrinsic axial symmetry-breaking at the needle tip means that the top and bottom surfaces of the platelike extension grow differently, often yielding one convex and one concave basal surface in the process. Although this seems to complicate matters, it often yields a cleaner analysis, as different morphological structures typically appear on the two basal surfaces [2021Lib]. Spatially separating the convex and concave growth phenomena generally makes it easier to examine both in detail.

The ice-needle method also eliminates the uncertainties associated with substrate interactions, and it is possible to achieve quite high supersaturations with no complications from droplet formation. Of course, the presence of the growing ice needle adds a nontrivial element to the analysis, but this can be incorporated into computational models. In many cases, the needle even provides a "witness surface" that can be used to sample and calibrate the far-away supersaturation in the experiment [2016Lib].

Comparing these three experimental strategies in detail, I have found that growing small prisms on hydrophobic substrates is an especially good technique for examining growth rates below $\sigma_{water}$, especially when it is desirable to do so at low background gas pressures. Substrate interactions remain a nagging source of uncertainty in these experiments, but this is usually a manageable problem. Levitation and free-fall methods are also quite useful below $\sigma_{water}$, but the technical issues are substantial compared with the simplicity of dropping ice prisms onto a flat substrate.

The use of ice needles as seed crystals has considerable advantages over the other strategies when the goal is to example full 3D morphological development over a broad range of growth conditions, especially at supersaturations at and above $\sigma_{water}$. I have recently made considerable progress exploring quantitative aspects of the molecular attachment kinetics using a simple witness-surface analysis of these crystals [2019Lib2, 2020Lib, 2020Lib1, 2020Lib2], and a close examination of plates growing on ice needles led to a model explaining why triangular snow crystals grow under certain conditions [2021Lib1]. While these experimental results already demonstrate the usefulness of using ice needles as seed crystals, the full benefits of this technique will become apparent when it becomes possible to thoroughly investigate and model the numerous complex morphological behaviors that appear during snow crystal growth on ice needles.

## ❄ Electric Ice Needles

I have already described the production of slender ice needle crystals using applied electric fields at some length [2014Lib1, 2021Lib], so I provide only a summary here. The essential phenomenon of electrified ice needle growth was discovered about six decades ago [1963Mas], followed much later by a theoretical model attributing needle formation to an electrically induced growth instability [1998Lib, 2002Lib]. Remarkably, the ice-needle tip radius can be as small as 100 nm during growth, yielding tip growth velocities above 150 μm/sec in extreme cases.

This laboratory curiosity developed into a productive research tool when my students and I discovered that trace chemical impurities could be added to promote ice needle growth along the crystalline c-axis [2002Lib]. Figure 15 shows a dual-diffusion-chamber apparatus designed to exploit this phenomenon by making c-axis ice needles in the first chamber and then transporting them into the second chamber for subsequent growth. The first diffusion chamber is optimized for needle production, while the second is optimized to produce well-defined growth conditions in air with a broadly adjustable temperature and supersaturation.

Thermally conducting stainless steel walls in the second diffusion chamber provide horizonal isotherms throughout the chamber that exhibit a simple linear temperature gradient in the vertical direction, as illustrated in Figure 16. In this stably stratified body of air, unperturbed by convective currents, it is possible to model the interior supersaturation as demonstrated in Figure 17. These models have been quantitatively tested by measuring the growth of simple ice needles using an analytical model for the growth of cylindrical crystals [2016Lib].

Figure 15: (below) A dual diffusion chamber apparatus for observing snow crystal growth on electric ice needles. Diffusion Chamber 1 (DC1, on the right) provides the necessary conditions for creating c-axis electric needles easily and quickly. The e-needles are then transported to Diffusion Chamber 2 (DC2, on the left), which provides a well-controlled environment that can achieve a broad range of temperatures and supersaturation levels. The inside height of DC2 is 10 centimeters [2014Lib1].

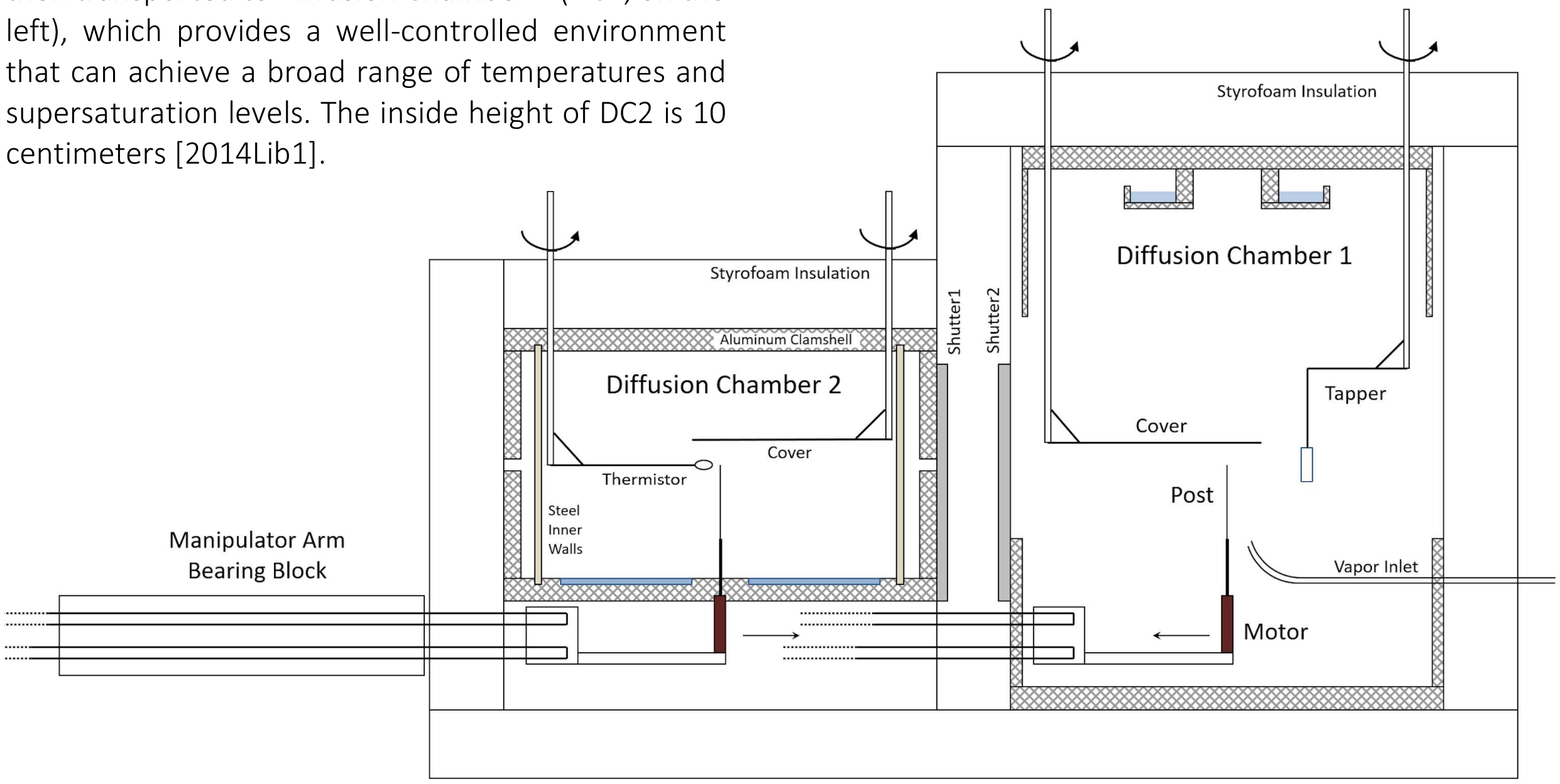

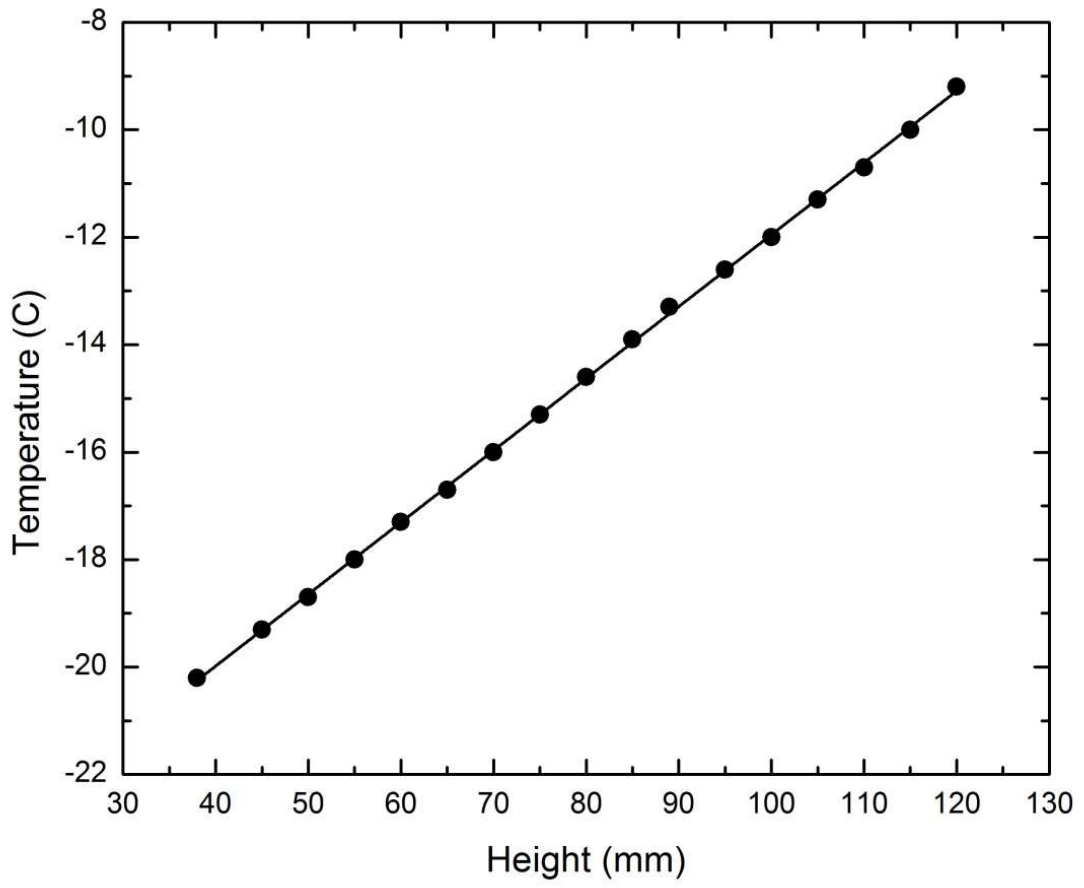

Figure 16: A typical vertical temperature profile in the second diffusion chamber, revealing a linear temperature profile that does not vary substantially with horizontal position within the chamber. This temperature profile ensures a stable, convection-free environment, allowing accurate models of the interior supersaturation as illustrated in Figure 17.

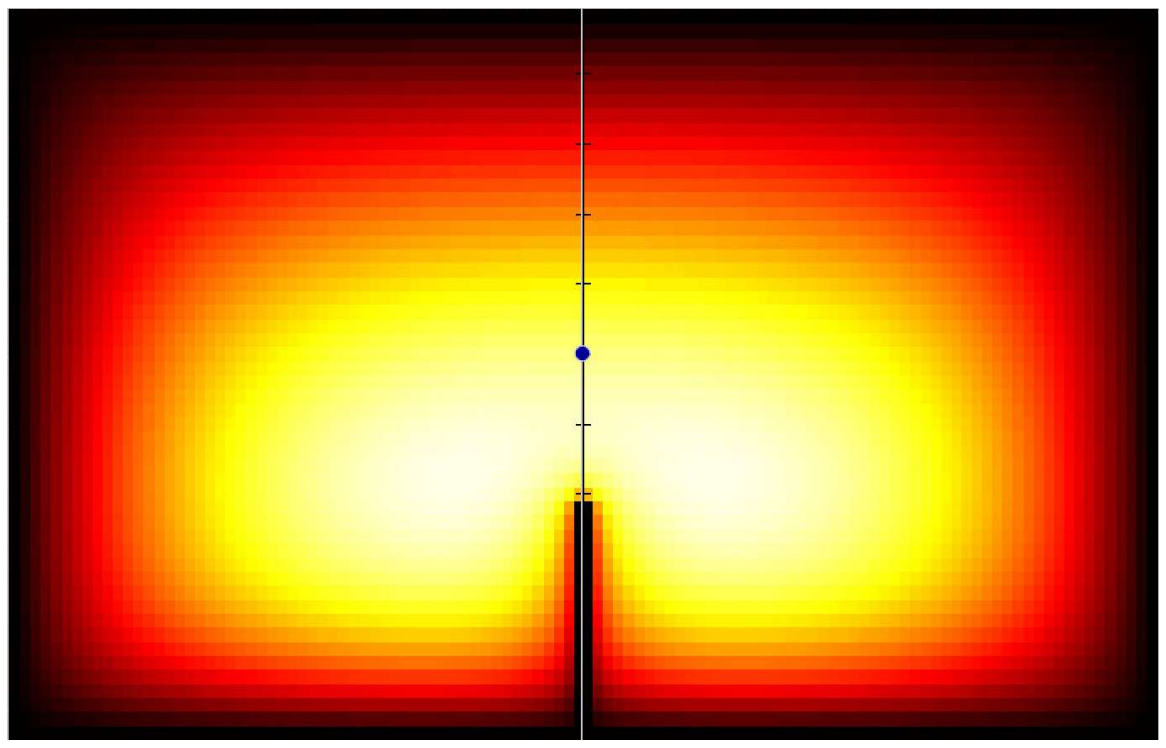

Figure 17: A typical model of the water-vapor supersaturation inside the second diffusion chamber. Note that the supersaturation goes to zero (black) at the ice-covered walls of the chamber. The central dot shows the position of crystals growing on the ends of c-axis ice needles.

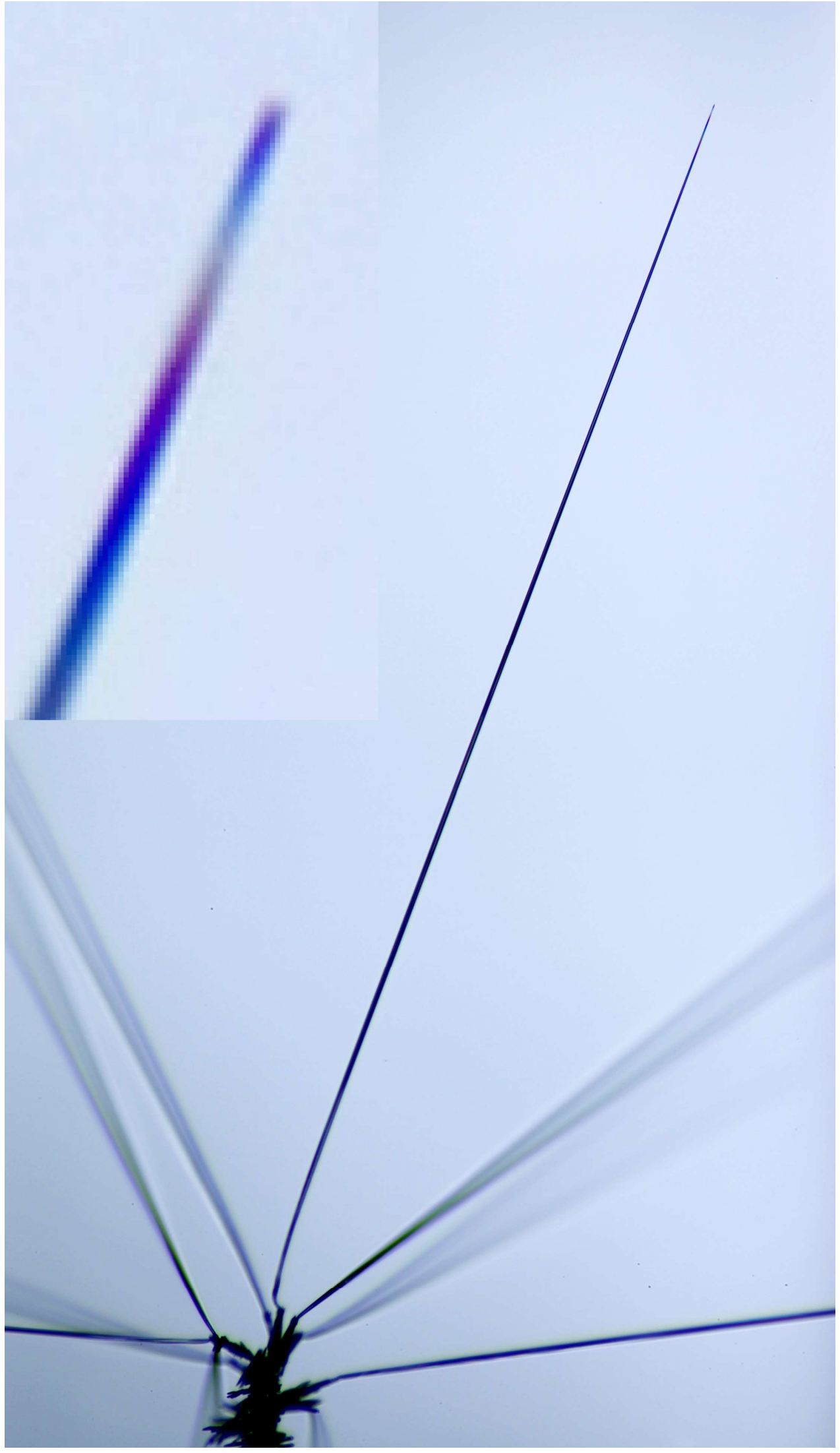

Figure 18: Several c-axis electric ice needles growing out from a frost-covered wire held at an electrical potential of 2000 volts. The longest needle is approximately 3 mm long, and the inset image shows a magnified view of the tip of the tallest needle. Diffraction effects can be seen near the needle tip, which has a tip radius of about 1.5 microns.

In operation, Figure 18 shows a typical set of c-axis electric ice needles growing on a frost-covered wire in the first diffusion chamber. Although the needles are quite slender with seemingly tenuous connections at the wire base, they are quite robust and usually survive the transport from the first to the second diffusion chamber. With the electric fields removed in the second chamber, normal growth commences from the needle tips, yielding a wide variety of growth morphologies depending on temperature and supersaturation. Figure 19 show one example with platelike crystals emerging near -15 C.

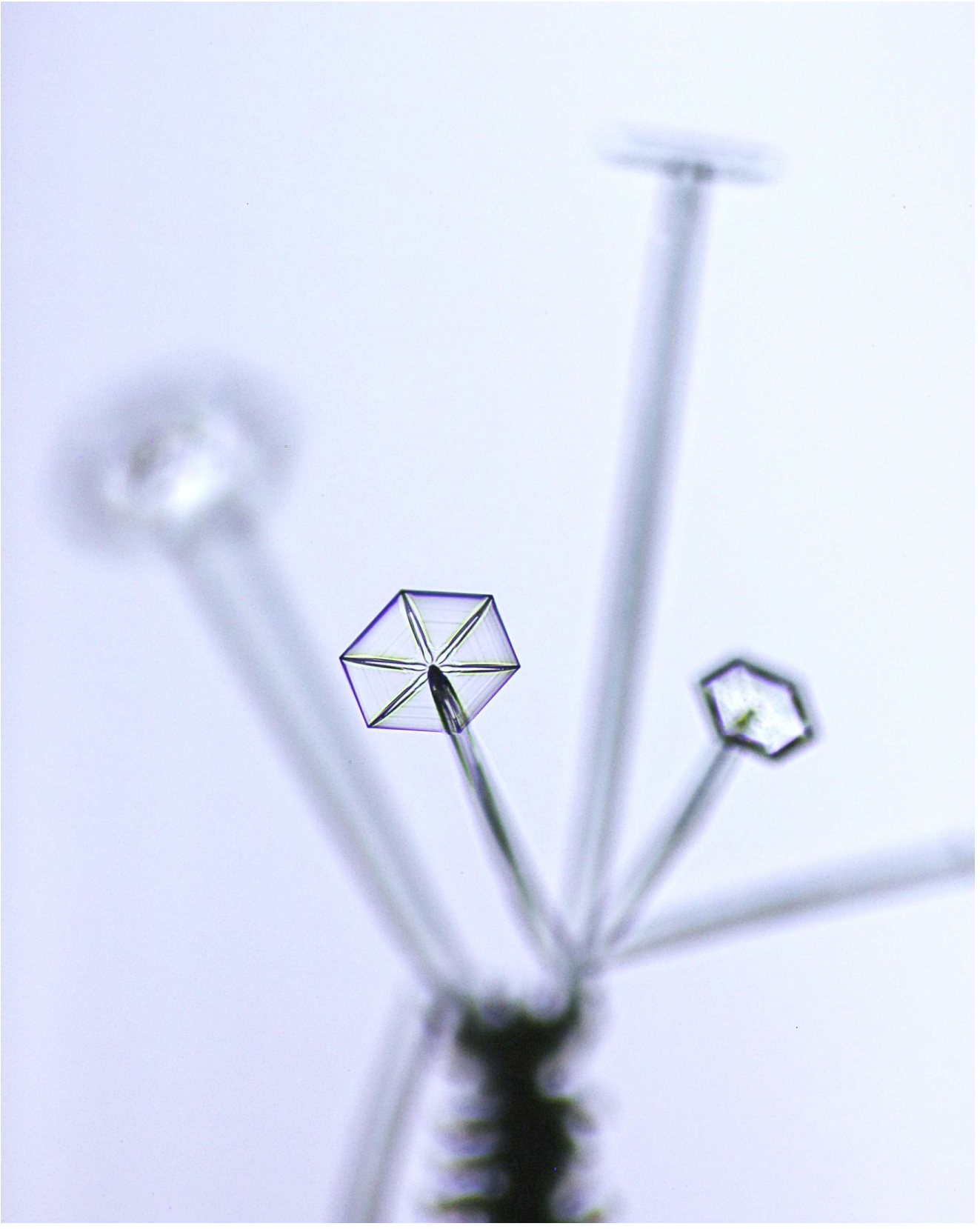

Figure 19: After this set of electric ice needles was transported to the second diffusion chamber and the electric fields removed, thin ice plates began growing from the needle tips.

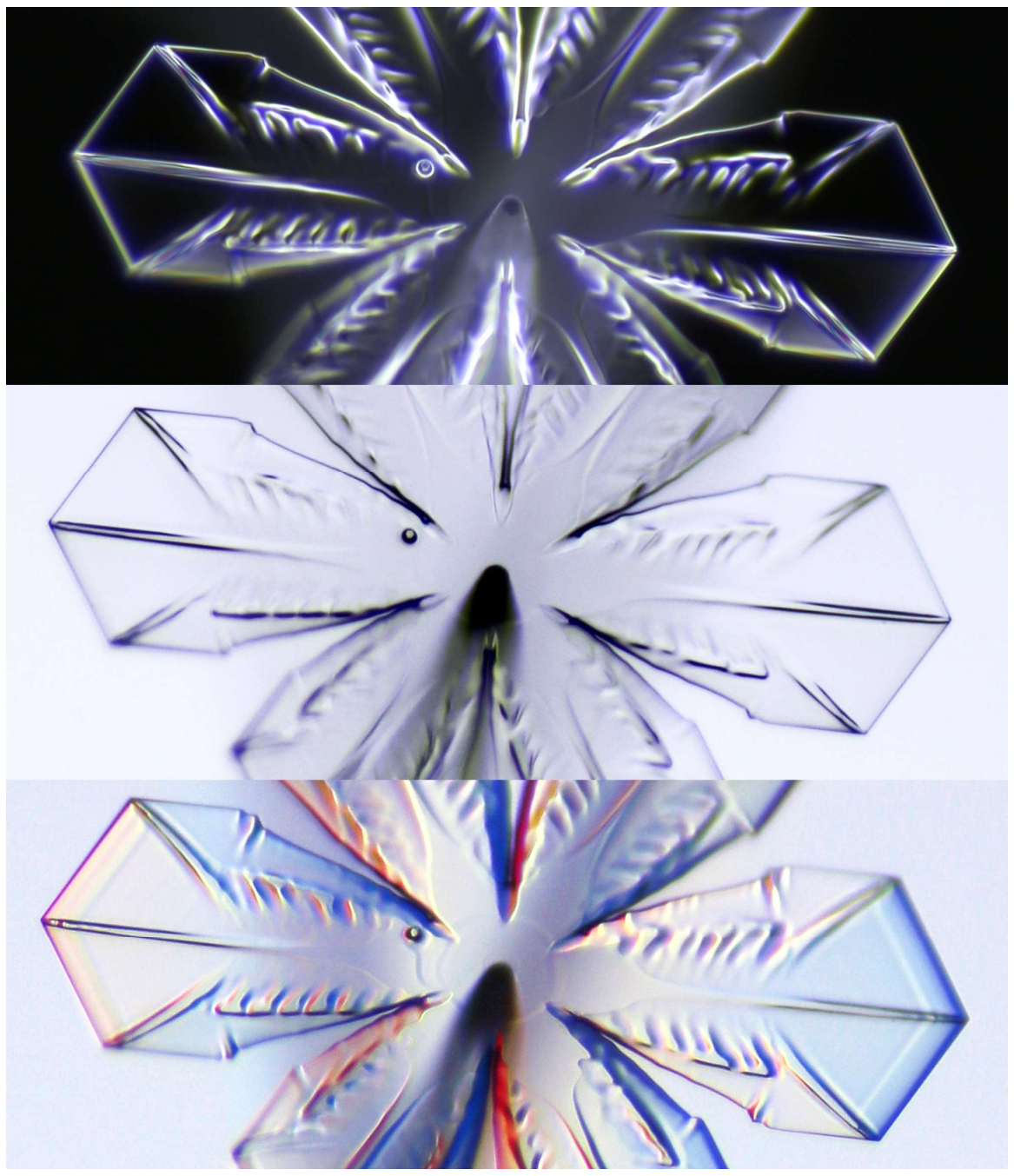

Figure 20: This series of photographs of the same growing crystal illustrates how different illumination types (all using transmitted light from behind the crystal) accentuate different structural features. Dark-field illumination (top) emphasizes edges, plain background illumination (center) gives the highest resolution, and Rheinberg illumination (bottom) yields high contrast on subtle surface structures.

## A Taxonomy of Snow Crystal Growth Behaviors

The potential for using the ice-needle method to make quantitative comparisons with computational models becomes apparent when observing growth over a broad range of environmental conditions. Figure 24 at the end of this paper shows what is essentially a pictorial Nakaya diagram of snow-crystal growth morphologies as a function of temperature and supersaturation in air. These images were obtained using constant growth environments in each case, but it is certainly possible to explore time-dependent conditions as well. Moreover, these images only illustrate qualitative snapshots of crystal morphologies under different growth conditions. Quantitative observations of the full growth history can easily be recorded as well for detailed comparisons with computational models.

During the process of obtaining these and other observations, it soon became apparent that a several morphological features are especially common, suggesting that they are robust with respect to small changes in growth conditions. As has been known for many decades, platelike forms regularly appear around -14 C and at temperatures above -3 C, while hollow columns emerge at temperatures around -5 C. A closer look, however, reveals that a variety of complex structures are included on these basic shapes, including some I have been calling ridges, fins, and I-beams [2021Lib]. Basic ridges are illustrated in Figure 20, while Figure 21 shows side fins on a conical cuplike crystal, and Figure 22 demonstrates an observation of I-beam structures under a slightly conical plate.

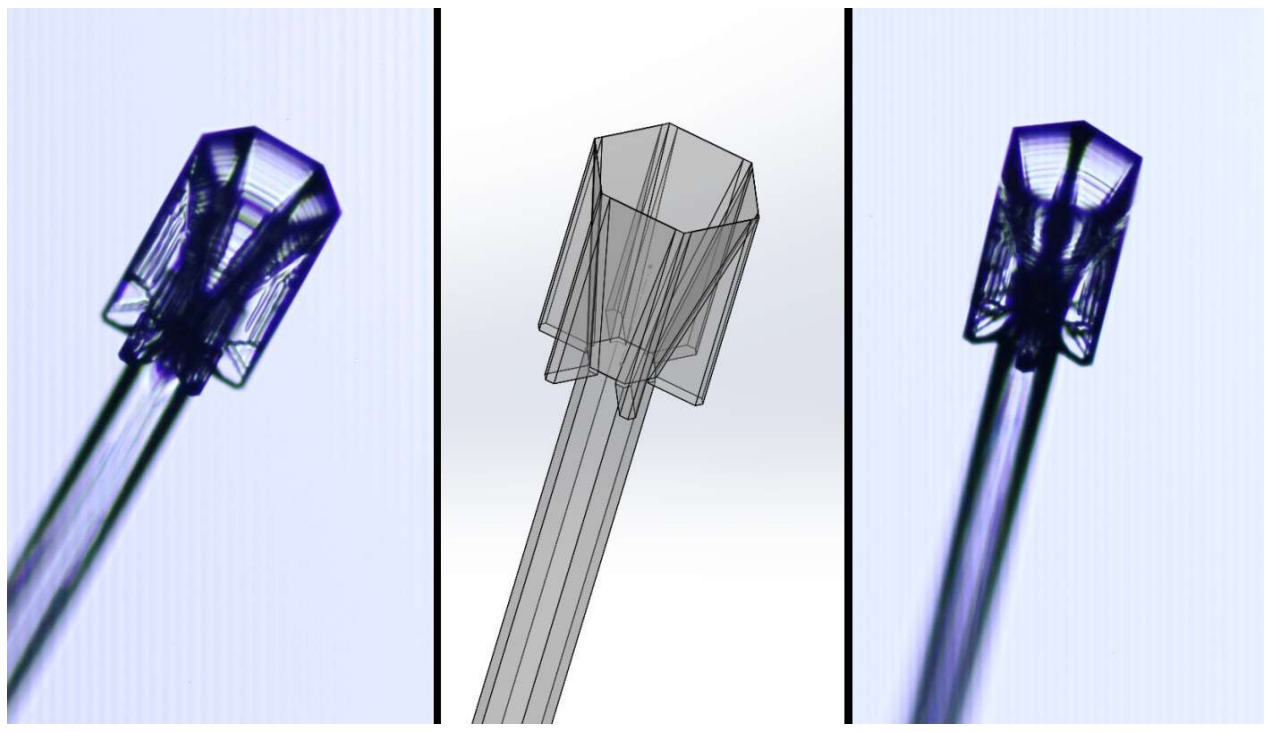

Figure 21: This hollow cuplike crystal formed in conditions near (T,σ) = (-7 C, 32%), flanked by six plate-like "fins" on the sides. The accompanying 3D sketch is meant to clarify the structure, which can be difficult to discern from photographs. As the temperature is lowered, the opening angle of the cup increases, eventually transforming this morphology into a slightly conical plate with I-beam ridges. [SolidWorks drawing by Ryan Potter.]

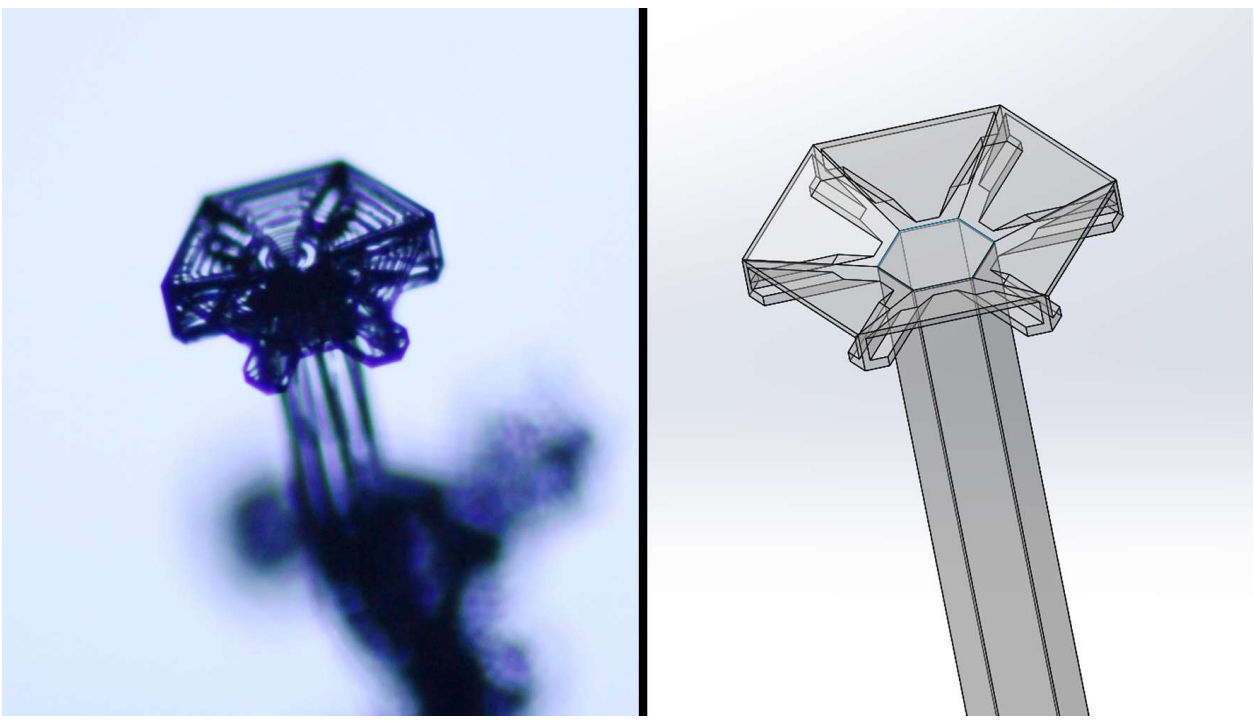

Figure 22: This fairly common e-needle morphology exhibits a slightly conical plate, pronounced ridges on the undersurface of the plate, and platelike extensions growing from the bottoms of the ridges. The ridge structure thus has a general "I-beam" appearance. As seen in Figure 24, variations of the I-beam ridge structure can be found in many regions of the e-needle morphology diagram. [SolidWorks drawing by Ryan Potter.]

As the supersaturation increases, snow crystal growth develops into highly complex dendritic structures, and Figure 24 shows how these vary with temperature. Although snow-crystal dendrites are often depicted as flat and essentially two-dimensional, these observations show how real ice/vapor dendrites exhibit a range of three-dimensional structures, including the "fishbone" dendrites shown in Figure 23 [2009Lib1]. Highly dendritic growth is inevitable at sufficiently high supersaturations but creating realistic computational models of these complex structures remains an ongoing challenge. Reproducing the full spectrum of snow crystal dendrites will require diffusion-limited growth together with a suitable parameterized model of the temperature-dependent, anisotropic attachment kinetics.

Having performed an initial survey of parameter space in my observations of snow crystal growth on ice needles thus far, it has become clear that the ice-needle technique offers a substantial new opportunity for improving our understanding of the essential physics underlying this fascinating phenomenon. The computational modeling algorithms are becoming sufficiently advanced that it is already possible to produce model crystals that recreate some of the basic structures shown in Figure 24. Making detailed comparisons between laboratory experiments and quantitative observations will doubtless lead to new insights in the ice/vapor attachment kinetics and the molecular processes needed to accurately model structure formation during faceted+branched crystal growth.

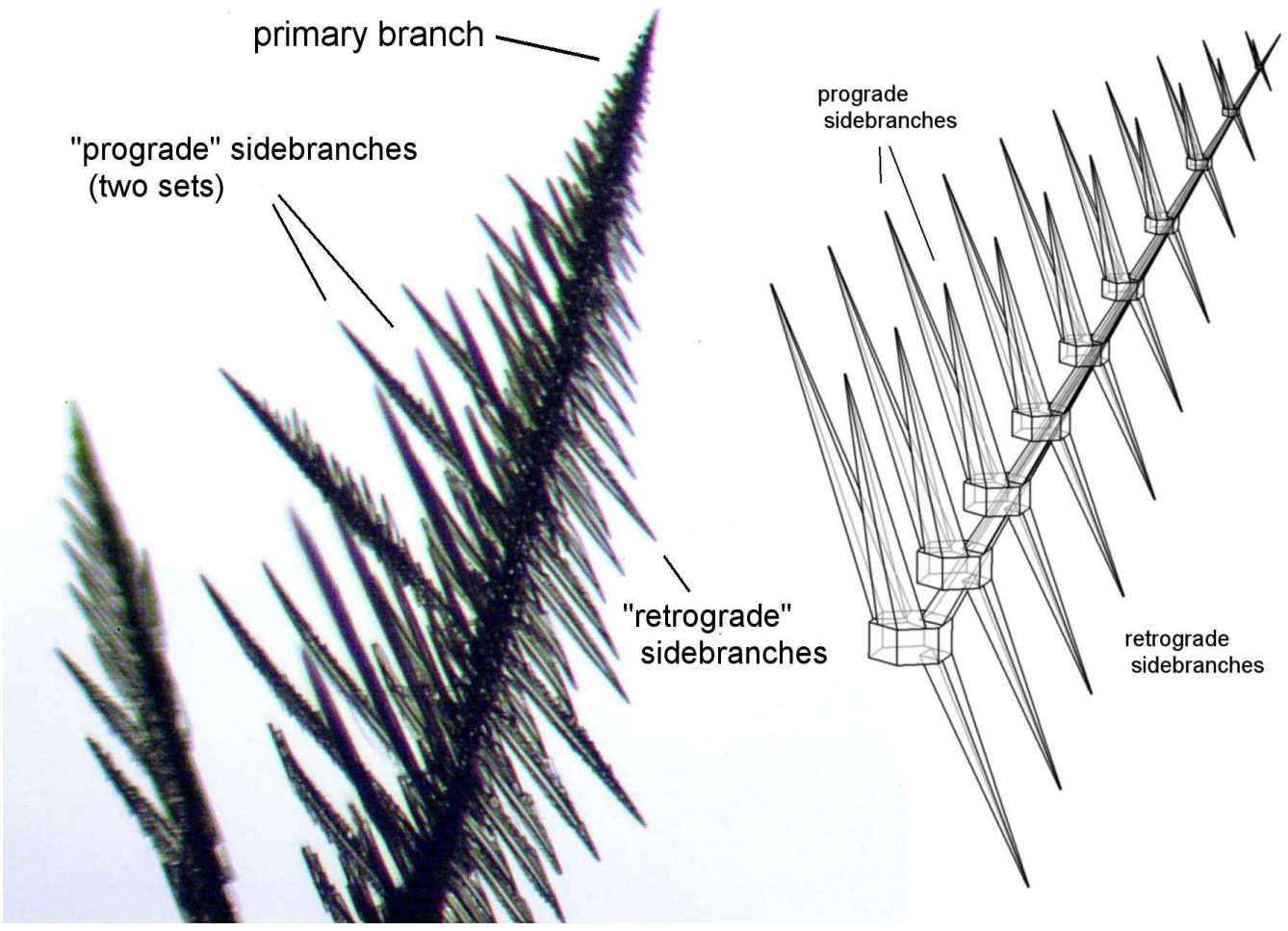

Figure 23: At temperatures near -5 C, snow crystal dendrites appear at high supersaturations with a "fishbone" structure consisting of both prograde and retrograde sidebranches. Reproducing the full range of temperature-dependent dendritic structures presents a significant challenge for computational models of snow crystal growth [2009Lib1].

The purpose of this planned series of papers is to further explore and develop a quantitative taxonomy of snow crystal growth behaviors over a broad range of growth conditions. My preliminary observations suggest that there is much to be learned by focusing detailed investigations on specific regions in temperature-supersaturation parameter space, examining specific growth morphologies and morphological transitions. My overarching goal is thus to take the Nakaya diagram to the next level of observational detail, providing quantitative experimental data that can be incorporated into future computational modeling efforts.

## ❄ References

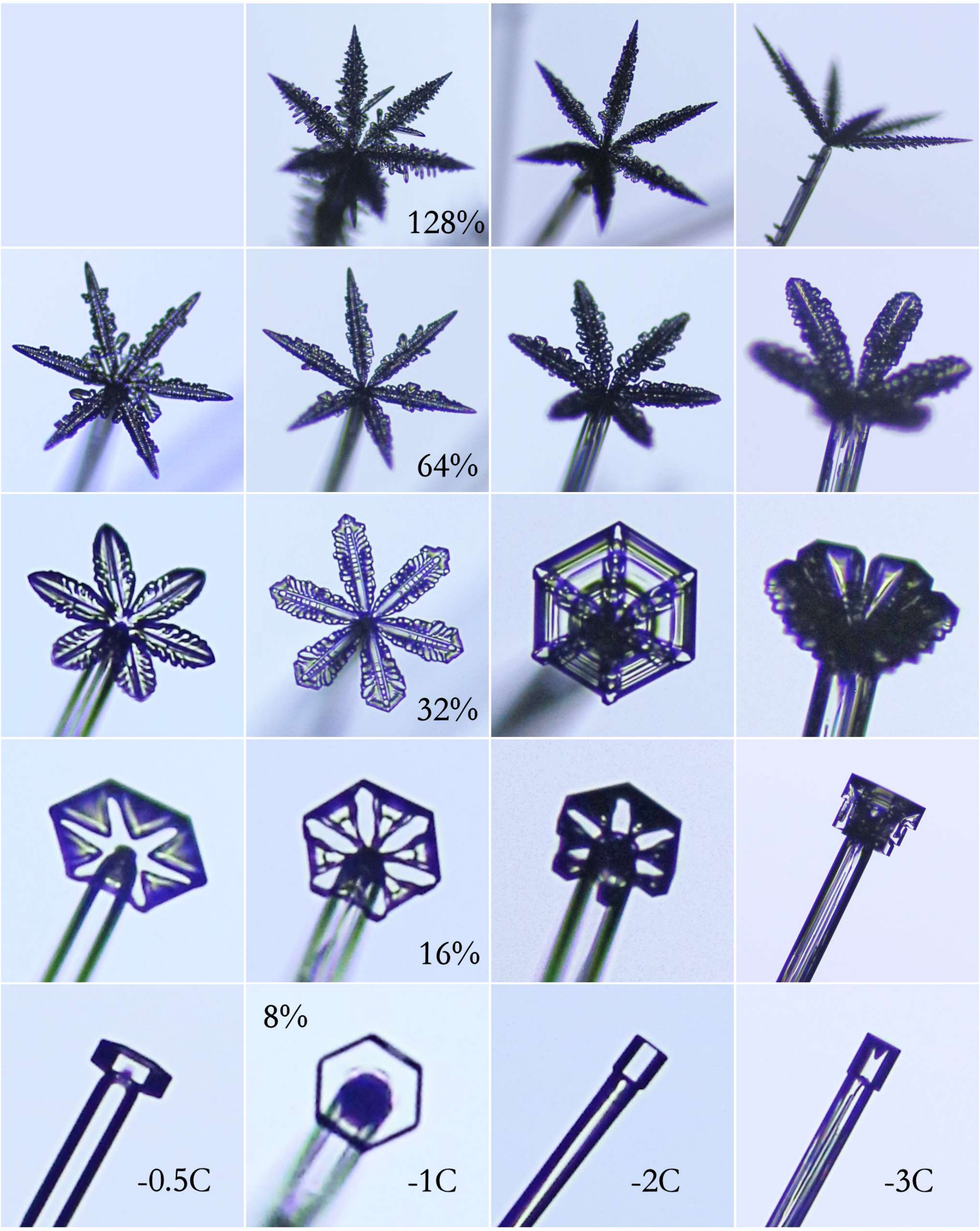

Figure 24a. The Nakaya diagram illustrated by snow crystals growing on the ends of slender ice needles in air. The horizontal and vertical axes indicate temperature and far-away supersaturation as labeled. Platelike growth is common at high temperatures, with greater morphological complexity as the supersaturation increases. Plates transition to somewhat blockier forms as the temperature falls from -0.5 C to -3 C. Dendritic sidebranching is weak in this temperature range, and the dendrite growth direction varies with temperature and supersaturation. Prism and basal faceting are both strong even at -0.5 C. The upper-left panel is missing because the fast growth rate at those conditions causes melting.

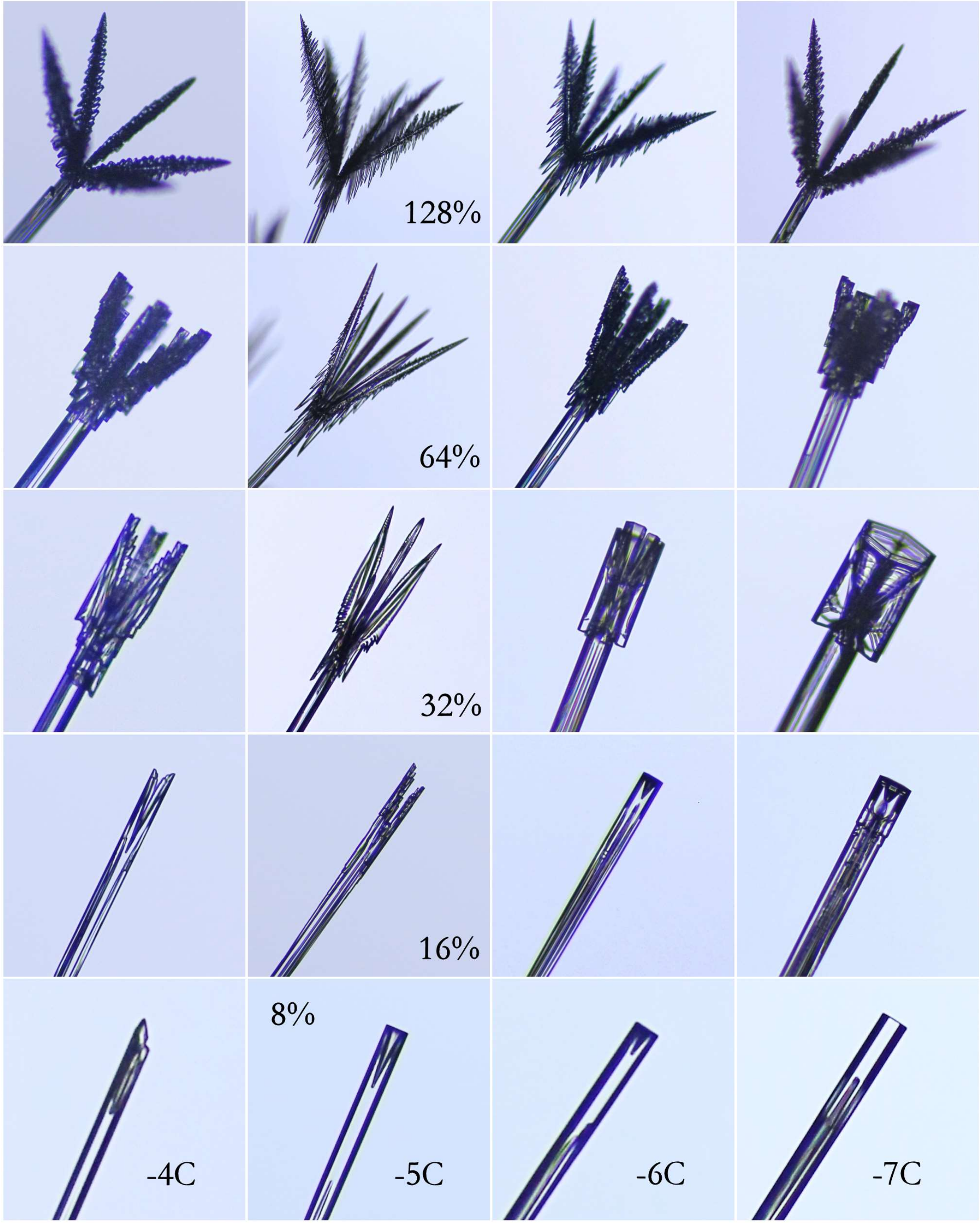

Figure 24b. The Nakaya diagram illustrated by snow crystals growing on the ends of slender ice needles in air. The horizontal and vertical axes indicate temperature and far-away supersaturation as labeled. Columns and needles are common near -5 C, turning into distinctive "fishbone" dendrites at the highest supersaturations. Hollow columns appear at (-5C, 8%), branching into needles at (-5C, 16%). Exceptionally thin-walled cups can be seen at (-7C, 32%). The "tridents" at (-5C, 32%) result from competition between neighboring branches (Chapter 3).

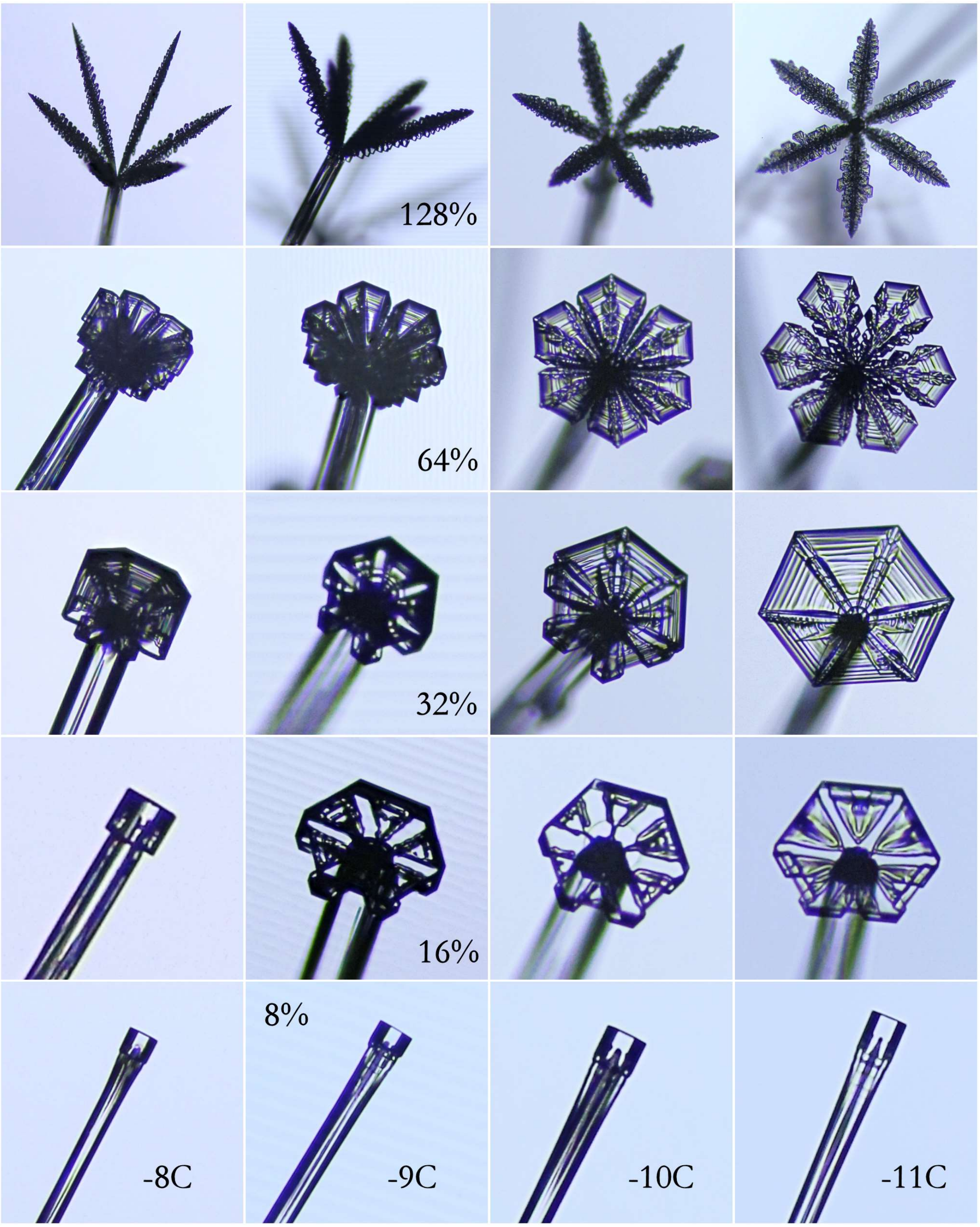

Figure 24c. The Nakaya diagram illustrated by snow crystals growing on the ends of slender ice needles in air. The horizontal and vertical axes indicate temperature and far-away supersaturation as labeled. The basal and prism growth rates are nearly identical at -8 C, yielding blocky forms and weak sidebranching at high supersaturations. Thin plates emerge as temperatures drop just a few degrees below -8 C. Deep ridging is seen at (-8C, 32%), developing into I-beam structures at -9 C and -10 C (Chapter 3). Ridges generally tend to become thinner as the temperature moves toward -15 C.

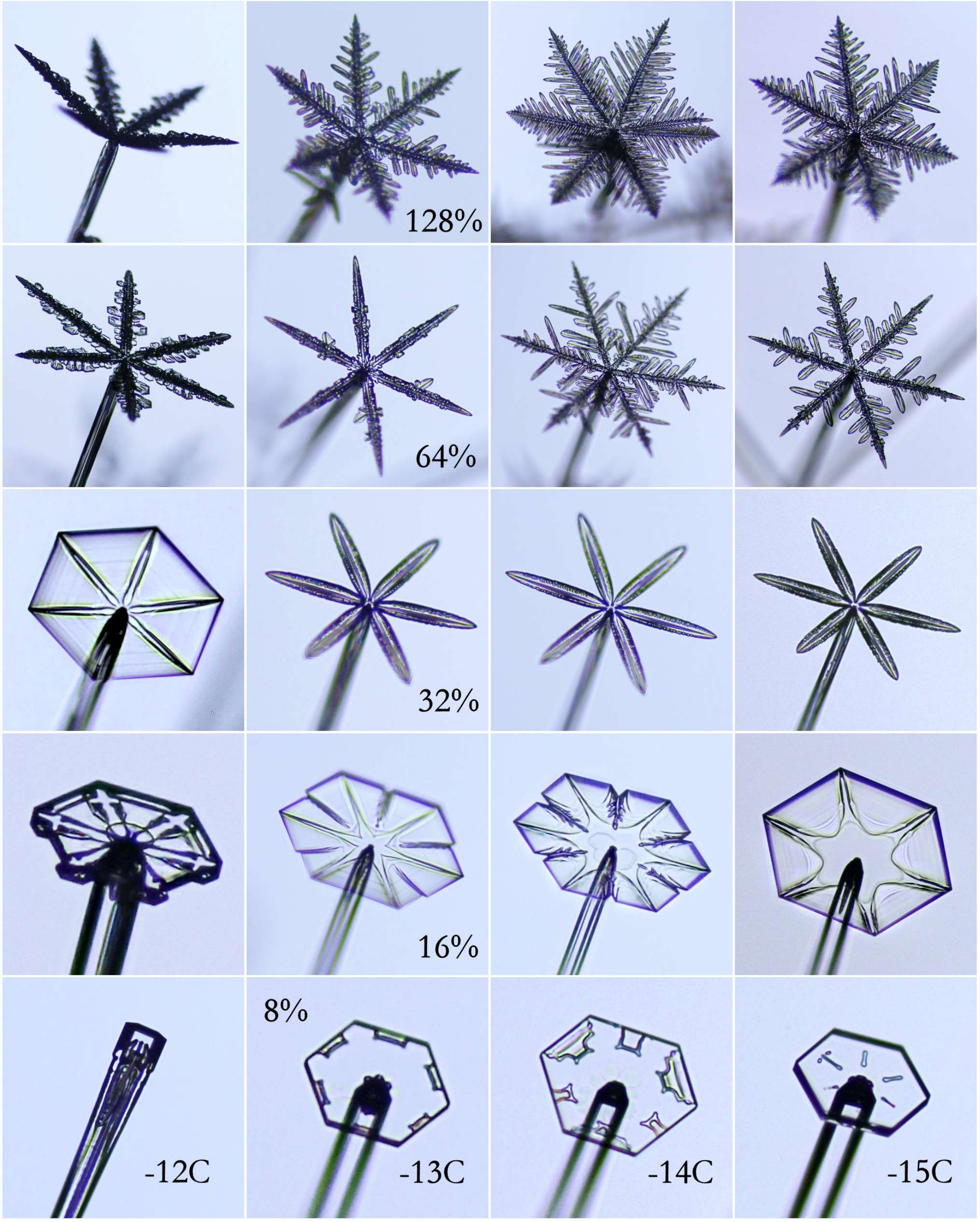

Figure 24d. The Nakaya diagram illustrated by snow crystals growing on the ends of slender ice needles in air. The horizontal and vertical axes indicate temperature and far-away supersaturation as labeled. Exceptionally thin plates appear in a narrow temperature range near -14 C, accompanied by nearly flat fernlike dendrites at high supersaturations, exhibiting exceptionally well developed sidebranching. Simple starts are common around (-14C, 32%), quickly transitioning to dendrites at higher supersaturations.

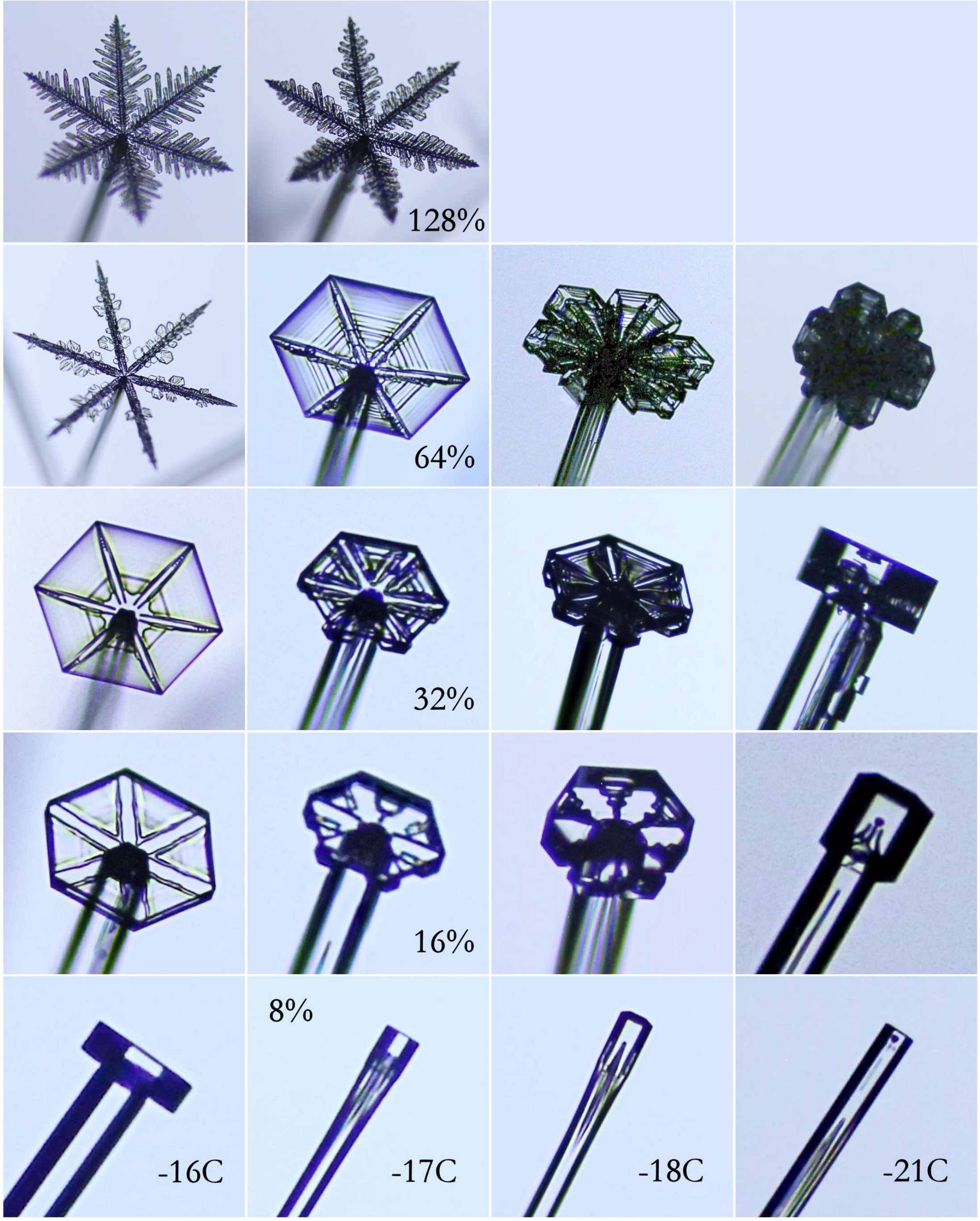

Figure 24e. The Nakaya diagram illustrated by snow crystals growing on the ends of slender ice needles in air. The horizontal and vertical axes indicate temperature and far-away supersaturation as labeled. Thin platelike crystals transition to blockier forms as the temperature drops from -15 C to -21 C, with I-beam structures appearing during the transition. Hollow plates form at (-16C, 16%), while much thinner plates appear at (-16C, 32%). Panels on the upper right are missing because it is difficult to reach high supersaturations at low temperatures in a linear diffusion chamber.